\documentclass[useAMS]{mn2e}
\usepackage{times}
\usepackage{psfig}

\def\b{\beta}

\newcommand{\ltsima} {$\; \buildrel < \over \sim \;$}
\newcommand{\gtsima} {$\; \buildrel > \over \sim \;$}
\newcommand{\lta} {\lower.5ex\hbox{\ltsima}}
\newcommand{\gta} {\lower.5ex\hbox{\gtsima}}

\title[The power of blazar jets] 
{The power of blazar jets}

\author[A. Celotti \& G. Ghisellini]
{Annalisa Celotti$^1$\thanks{E--mail: celotti@sissa.it} 
and Gabriele Ghisellini$^2$ \\
$^1$ S.I.S.S.A., V. Beirut 2--4, I-34014 Trieste, Italy\\
$^2$ INAF -- Osservatorio Astronomico di Brera, V. Bianchi 46, 
I--23807 Merate, Italy}

\begin{document}

\pagerange{\pageref{firstpage}--\pageref{lastpage}} \pubyear{2002}
\maketitle

\begin{abstract}
  We estimate the power of relativistic, extragalactic jets by
  modelling the spectral energy distribution of a large number of
  blazars.  We adopt a simple one--zone, homogeneous, leptonic
  synchrotron and inverse Compton model, taking into account seed
  photons originating both locally in the jet and externally.  The
  blazars under study have an often dominant high energy component,
  which, if interpreted as due to inverse Compton radiation, limits the
  value of the magnetic field within the emission region.  As a
  consequence, the corresponding Poynting flux cannot be energetically
  dominant.  Also the bulk kinetic power in relativistic leptons is
  often smaller than the dissipated luminosity.  This suggests that
  the typical jet should comprise an energetically dominant proton
  component.  If there is one proton per relativistic electrons, jets
  radiate around 2--10 per cent of their power in high power blazars
  and 3--30 per cent in less powerful BL Lacs.
\end{abstract}

\section{Introduction}

The radiation observed from blazars is dominated by the emission from
relativistic jets (Blandford \& Rees 1978) which transport energy and
momentum to large scales. As the energy content on such scales already
implies in some sources jet powers comparable with that which can be
produced by the central engine (e.g. Rawlings \& Saunders 1991), only
a relatively small fraction of it can be radiatively dissipated on the
`blazar' (inner) scales.

However we still do not know the actual power budget in jets and in
which form such energy is transported, namely whether it is mostly
ordered kinetic energy of the plasma and/or Poynting flux.  In
addition, the predominance of one or the other form can change during
their propagation. These of course are crucial pieces of information
for the understanding on how jets are formed and for quantifying the
energy deposition on large scales.

In principle the observed radiation can -- via the modelling of the
radiative dissipation mechanism -- set constraints on the minimum jet
power and can even lead to estimates of the relative contribution of
particles (and the corresponding bulk kinetic power), radiation and
magnetic fields.  The modelling depends of course on the available
spectral information and conditions on the various jet scales
(i.e. distances from the central power--house).  Attempts in this
direction include the work by Rawlings \& Saunders (1991), who
considered the energy contained in the extended radio lobes of
radio--galaxies and radio--loud quasars.  By estimating their
life--times they could calculate the average power needed to sustain
the emission from the lobe themselves (Burbidge 1959).

At the scale of hundreds of kpc the {\it Chandra} satellite
observations, if interpreted as inverse Compton scattering on the
Cosmic Microwave Background (Tavecchio et al. 2000; Celotti,
Ghisellini \& Chiaberge 2001), indicate that jets of powerful blazars
are still relativistic.  This allowed Ghisellini \& Celotti (2001) to
estimate a minimum power at these distances for PKS 0637--752, the
first source whose large scale X--ray jet was detected by {\it
  Chandra}.  Several other blazars were studied by Tavecchio et
al. (2004), Sambruna et al. (2006) and Tavecchio et al. (2007) who
found that the estimated powers at large scales were comparable
(within factors of order unity) with those inferred at much smaller
blazar scales.

Celotti \& Fabian (1993) considered the core of jets, as observed by
VLBI techniques, to derive a size of the emitting volume and the
number of emitting electrons needed to account for the observed radio
luminosity.  The bulk Lorentz factor, which affects the quantitative
modelling, was estimated from the relativistic beaming factor required
not to overproduce, by synchrotron self--Compton emission, the
observed X--ray flux (e.g. Celotti 1997).

A great advance in our understanding of blazars came however with the
discovery that they are powerful $\gamma$--ray emitters (Hartman et
al. 1999).  Their $\gamma$--ray luminosity often dominates (in the
powerful flat spectrum radio--loud quasars, FSRQs) the radiative
power, and its variability implies a compact emitting region.  The
better determined overall spectral energy distribution (SED) and total
observed luminosity of blazars constrain -- via pair opacity arguments
(Ghisellini \& Madau 1996) -- the location in the jet where most of
the dissipation occurs.  For a given radiation mechanism the modelling
of the SED also allows to estimate the power requirements and the
physical conditions of this emitting region.  Currently the models
proposed to interpret the emission in blazars fall into two broad
classes.  The so-called `hadronic' models invoke the presence of
highly relativistic protons, directly emitting via synchrotron or
inducing electron--positron (e$^{\pm}$) pair cascades following
proton-proton or proton--photon interactions (e.g., Mannheim 1993,
Aharonian 2000, M{\"u}cke et al. 2003, Atoyan \& Dermer 2003).  The
alternative class of models assumes the direct emission from
relativistic electrons or e$^{\pm}$ pairs, radiating via the
synchrotron and inverse Compton mechanism. Different scenarios are
mainly characterised by the different nature of the bulk of the seed
photons which are Compton scattered. These photons can be produced
both locally via the synchrotron process (SSC models, Maraschi,
Ghisellini \& Celotti 2002), and outside the jet (External Compton
models, EC) by e.g. the gas clouds within the Broad Line Region (BLR;
Sikora, Begelman \& Rees 1994, Sikora et al. 1997) reprocessing
$\sim$10 per cent of the disk luminosity.  Other contributions may
comprise synchrotron radiation scattered back by free electrons in the
BLR and/or around the walls of the jet (mirror models, Ghisellini \&
Madau 1996), and radiation directly from the accretion disk (Dermer \&
Schlickeiser 1993; Celotti, Ghisellini \& Fabian, 2007).

Some problems suffered by hadronic scenarios (such as pair
reprocessing, Ghisellini 2004) make us favour the latter class of
models. By reproducing the broad band properties of a sample of
$\gamma$--ray emitting blazars via the SSC and EC mechanisms, Fossati
et al. (1998) and Ghisellini et al. (1998; hereafter G98) constrained
the physical parameters of a (homogeneous) emitting source.  A few
interesting clues emerged.  The luminosity and SED of the sources
appear to be connected, and a spectral sequence in which the energy of
the two spectral components and the relative intensity decrease with
source power seems to characterise blazars, from low power BL Lacs to
powerful FSRQs (opposite claims have been put forward by Giommi et
al. 2007; see also Padovani 2006).
This SED sequence translates into an (inverse) correlation between the
energy of particles emitting at the spectral peaks and the energy
density in magnetic and radiation fields (Ghisellini, Celotti \&
Costamante 2002; hereafter G02).  An interpretation of such findings
is possible within the internal shock scenario (Ghisellini 1999; Spada
et al. 2001; Guetta et al 2004), which could account for the radiative
efficiency, location of the dissipative region and spectral trend if
the particle acceleration process is balanced by the radiative
cooling. In such a scenario the energetics on scales of 10$^2-10^3$
Schwarzschild radii is dominated by the power associated to the bulk
motion of plasma. This is in contrast with an electromagnetically
dominated flow (Blandford 2002; Lyutikov \& Blandford 2003).

Within the frame of the same SSC and EC emission model, in this work
we consider the implications on the jet energetics, the form in which
the energy is transported and possibly the plasma composition.  In
particular we estimate the (minimum) power which is carried by the
emitting plasma in electromagnetic and kinetic form in a significant
sample of blazars at the scale where the $\gamma$--ray emission -- and
hence most of the luminosity -- is produced.  Such scale corresponds
to a distance from the black hole of the order of $10^{17}$ cm
(Ghisellini \& Madau 1996), a factor 10--100 smaller than the VLBI
one.  The found energetics are lower limits as they only consider the
particles required to produce the observed radiation, and neglect
(cold) electrons not contributing to the emission.

In Section 2 the sample of sources is presented. In Section 3 we
describe how the power in particles and field have been estimated, and
the main assumptions of the radiative model adopted.  The results are
reported in Section 4 and discussed in Section 5.  Preliminary and
partial results concerning the power of blazar jets were presented in
conference proceedings (see e.g. Ghisellini 1999, 
Celotti 2001, 
Ghisellini 2004). 

We adopt a concordance cosmology with $H_0=70$ km s$^{-1}$ Mpc$^{-1}$,
$\Omega_\Lambda=0.7$ and $\Omega_{\rm M}=0.3$.

\section{The sample}

The sample comprises the blazars studied by G98, namely all blazars
detected by EGRET or in the TeV band (at that time) for which there is
information on the redshift and on the spectral slope in the
$\gamma$--ray band.

To those, FSRQs identified as EGRET sources since 1998 or not present
in G98 have been added, namely:
PKS 0336--019 (Mattox et al. 2001); Q0906+6930 (the most distant
blazar known, at $z=5.47$, Romani et al. 2004; Romani 2006); PKS
1334--127 (Hartman et al. 1999; modelled by Foschini et al. 2006); PKS
1830--211 (Mattox et al. 1997; studied and modelled by Foschini et
al. 2006); PKS 2255--282 (Bertsch 1998; Macomb, Gehrels \& Shrader
1999).
and the three high redshift ($z>4$) blazars 
0525--3343; 1428+4217 and 1508+5714,
discussed and modelled in G02.

As for BL Lacs, we have included 0851+202 (identified as an EGRET
source, Hartman et al. 1999; modelled by Costamante \& Ghisellini
2002; hereafter C02) and those detected in the TeV band besides Mkn
421, Mkn 501, and 2344+512, which were already present in G98.  These
additional TeV BL Lacs are:
1011+496 (Albert et al. 2007c; see C02); 
1101--232 (Aharonian et al. 2006a; see C02 and G02); 
1133+704 (Albert et al. 2006a; see C02);
1218+304 (Albert et al. 2006b; see C02 and G02); 
1426+428 (Aharonian et al. 2002; Aharonian et al. 2003; see G02); 
1553+113 (Aharonian et al. 2006b; Albert et al. 2007a; see C02); 
1959+650 (Albert et al. 2006c; see C02); 
2005--489 (Aharonian et al. 2005a; see C02 and G02); 
2155--304 (Aharonian et al. 2005b; Aharonian et al. 2007b; 
already present in G98 as an EGRET source); 
2200+420 (Albert et al. 2007b; already present in G98 as an EGRET source); 
2356--309 (Aharonian et al. 2006a; Aharonian et al. 2006c; see C02 and G02).
Finally, we have considered the BL Lacs modelled in G02, namely
0033+505, 0120+340, 0548--322 and 1114+203.

In all cases the observational data were good enough to determine the
location of the high energy peak, a crucial information to constrain
the model input parameters.  The total number of sources is 74: 46
FSRQs and 28 BL Lac objects, 14 of which are TeV detected sources.
The objects are listed in Table 1 together with the input parameters
of the model fit.

\section{Jet powers: assumptions and method}

As already mentioned and widely assumed, the infrared to $\gamma$--ray
SED of these sources was interpreted in terms of a one--zone
homogeneous model in which a single relativistic lepton population
produces the low energy spectral component via the synchrotron process
and the high energy one via the inverse Compton mechanism.  Target
photons for the inverse Compton scattering comprise both synchrotron
photons produced internally to the emitting region itself and photons
produced by an external source, whose spectrum is represented by a
diluted blackbody peaking at a (comoving) frequency $\nu^\prime\sim
10^{15}\Gamma$ Hz.  We refer to G02 for further details about the
model.

The emitting plasma is moving with velocity $\beta c$ and bulk Lorentz
factor $\Gamma$, at an angle $\theta$ with respect to the line of
sight. The observed radiation is postulated to originate in a zone of
the jet, described as a cylinder, with thickness $\Delta R'\sim R$ 
as seen in the comoving frame, and volume $\pi R^2 \Delta
R^\prime$.  $R$ is the cross section radius of the jet.

The emitting region contains the relativistic emitting leptons and
(possibly) protons of comoving density $n_{\rm e}$ and $n_{\rm p}$,
respectively, embedded in a magnetic field of component $B$
perpendicular to the direction of motion, homogeneous and tangled
throughout the region. The model fitting allows to infer the physical
parameters of the emitting region, namely its size and beaming factor,
and of the emitting plasma, i.e. $n_{\rm e}$ and $B$.  These
quantities translate into jet kinetic powers and Poynting flux.

Assuming one proton per relativistic emitting electron and protons
`cold' in the comoving frame, the proton kinetic power corresponds to
\begin{equation}
L_{\rm p}\, \simeq \, \pi R^2 \Gamma^2 \b c \, n_{\rm p} m_{\rm p}
c^2,
\end{equation}
while relativistic leptons contribute to the kinetic power as:
\begin{equation}
L_{\rm e}\, \simeq \, \pi R^2 \Gamma^2 \b c \, n_{\rm e}\,
\langle\gamma\rangle m_{\rm e} c^2,
\end{equation}
where $\langle\gamma\rangle$ is the average random Lorentz factor of
the leptons, measured in the comoving frame, and $m_{\rm p}$, 
$m_{\rm e}$ are the proton and electron rest masses, respectively.

The power carried as Poynting flux is given by
\begin{equation}
L_{\rm B}\, \simeq \, {1 \over 8} R^2 \Gamma^2 \beta c B^2. 
\end{equation}

The observed synchrotron and self--Compton luminosities $L$ are
  related to the comoving luminosities $L^\prime$ (assumed to be
  isotropic in this frame) by $ L =\delta^4 L^\prime$, where the
  relativistic Doppler factor
  $\delta=[\Gamma(1-\beta\cos\theta)]^{-1}$.  The EC luminosity,
  instead, has a different dependence on $\theta$, being anisotropic
  in the comoving frame, with a boosting factor $\delta^6/\Gamma^2$
  (Dermer 1995). The latter coincides to that of the synchrotron and
  self--Compton radiation for $\delta=\Gamma$ i.e. when the viewing
  angle is $\theta \sim 1/\Gamma$.  For simplicity, we adopt a
  $\delta^4$ boosting for all emission components.

  Besides the jet powers corresponding to protons, leptons and
  magnetic field flowing in the jet, there is also an analogous
  component associated to radiation, corresponding to:
\begin{equation}
L_{\rm r} \, \simeq \, \pi R^2 \Gamma^2 \b c \, U^\prime_{\rm r}
\, \simeq L^\prime \Gamma^2
\end{equation}
where $U^\prime_{\rm r} = L^\prime/(\pi R^2 c)$ 
is the radiation energy density measured in the comoving frame.

%
%

We refer to G02 for a detailed discussion on the general robustness
and uniqueness of the values which are inferred from the
modelling. Here we only briefly recall the main assumptions of this
approach.

The relativistic particles are assumed to be injected throughout the
emitting volume for a finite time $t^\prime_{\rm inj}=\Delta R'/c$.
Since blazars are variable (flaring) sources, a reasonably good
representation of the observed spectrum can be obtained by considering
the particle distribution at the end of the injection, at
$t=t^\prime_{\rm inj}$, when the emitted luminosity is maximised.  In
this respect therefore the powers estimated refer to {\it flaring
states} of the considered blazars and do not necessarily represent
average values.

As the injection lasts for a finite timescale, only the higher energy
particles have time to cool (i.e. $t_{\rm c}< t_{\rm inj}$). The
particle distribution $N(\gamma)$ can be described as a broken
power--law with the injection slope below $\gamma_{\rm c}$ and steeper
above it.  We adopt a particle distribution $N(\gamma)$ that
corresponds to injecting a broken power--law with slopes $\propto
\gamma^{-1}$ and $\propto \gamma^{-s}$ below and above the break at
$\gamma_{\rm inj}$.  Thus the resulting shape of $N(\gamma)$ depends
on 1) the injected distribution and 2) the cooling time with respect
to $t_{\rm inj}$.

The limiting cases in relation to 2) can be identified with powerful
FSRQs and low power BL Lacs.  For FSRQs the cooling time is shorter
than $t_{\rm inj}$ for all particle energies (fast cooling regime) and
therefore the resulting $N(\gamma)$ is a broken power--law with a
break at $\gamma_{\rm inj}$, the energy of the leptons emitting most
of the observed radiation, i.e.
\begin{eqnarray}
N(\gamma) & \propto &\gamma^{-(s+1)}; \qquad \gamma > \gamma_{\rm inj}
\nonumber\\
N(\gamma) & \propto &\gamma^{-2}; \qquad \quad\,\,\,\, \gamma_{\rm c} < \gamma < \gamma_{\rm inj}
\nonumber\\
N(\gamma) & \propto &\gamma^{-1}; \qquad \quad\,\,\,\,  \gamma < \gamma_{\rm c}
\end{eqnarray}
For low power BL Lacs only the highest energy leptons can cool in
$t_{\rm inj}$ (slow cooling regime), and if the cooling energy (in
$t_{\rm inj}$) is $\gamma_{\rm inj}< \gamma_{\rm c} < \gamma_{\rm
max}$ ($\gamma_{\rm max}$ is the highest energy of the injected
leptons), we have:
\begin{eqnarray}
N(\gamma) &\propto& \gamma^{-(s+1)}; \qquad \,\,\, 
\gamma> \gamma_{\rm c} \nonumber\\
N(\gamma) &\propto& \gamma^{-s};\qquad \qquad\gamma_{\rm inj} <\gamma <
\gamma_{\rm c}\nonumber\\
N(\gamma) &\propto& \gamma^{-1};\qquad \qquad\gamma < \gamma_{\rm inj}.
\end{eqnarray}
For intermediate cases the detailed $N(\gamma)$ is fully described in G02.

\subsection{Dependence of the jet power on the assumptions}

We examine here the influence of the most crucial assumptions on the
estimated powers.

\begin{itemize}

\item {\bf Low energy cutoff ---} A well known crucial parameter for
  the estimates of powers in particles, which is poorly fixed by the
  modelling, is the low energy distribution of the emitting leptons,
  parametrised via a minimum $\gamma_{\rm min}$ (i.e. for say
  $\gamma<10$).  Indeed particles of such low energies (if present)
  would not contribute to the observed synchrotron spectrum, since
  they emit self--absorbed radiation.  They would instead contribute
  to the low energy part of the inverse Compton spectrum, but: i) in
  the case of SSC emission, their contribution is dominated by the
  synchrotron luminosity of higher energy leptons; ii) in the case of
  EC emission their radiation could be masked by the SSC (again
  produced by higher energy leptons) or by contributions from other
  parts of the jet.

  However in very powerful sources there are indications that the EC
  emission dominates in the X--ray range and thus the observations
  provide an upper limit to $\gamma_{\rm min}$.  In such sources there
  is direct spectral evidence that $\gamma_{\rm min}$ is close to
  unity.  Fig. \ref{1127} illustrates this point.  It can be seen how
  the model changes by assuming different $\gamma_{\rm min}$: only
  when $\gamma_{\rm min} \sim$1 a good fit of the soft X--ray spectrum
  can be obtained.  For such powerful blazars, the cooling time is
  short for leptons of all energies, ensuring that $N(\gamma)$ extends
  down at least to $\gamma_{\rm c}\sim$ a few.  The extrapolation of
  the distribution down to $\gamma_{\rm min}=1$ with a slope
  $\gamma^{-1}$ therefore implies that the possible associated error
  in calculating the number of leptons is $\ln(\gamma_{\rm c})$.

For low power BL Lacs the value of $\gamma_{\rm min}$ is much more
uncertain.  In the majority of cases $\gamma_{\rm c}>\gamma_{\rm
inj}$, and our extrapolation assuming again a $\gamma^{-1}$ slope
translates in an uncertainty in the lepton number $\sim
\ln(\gamma_{\rm inj})$.  Thus $L_{\rm e}$ and $L_{\rm p}$ could be
{\it smaller} up to this factor.

\begin{figure}
\vskip -0.5true cm
\hskip -2. true cm
\psfig{file=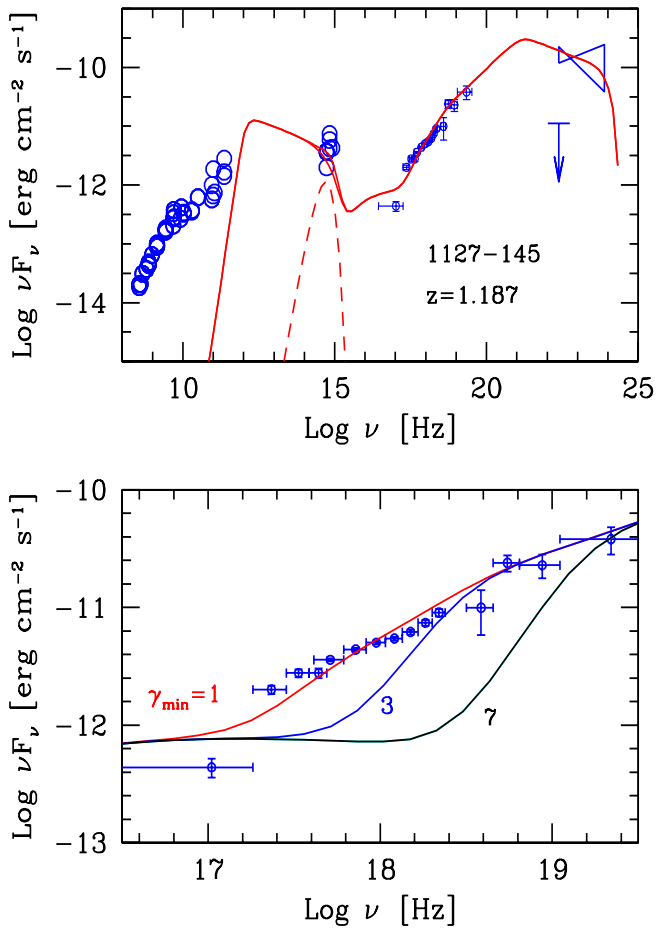,width=12.5truecm,height=14truecm}
\vskip -0.7true cm
\caption{Top panel: SED of 1127--145 `fitted' by our model, assuming
  $\gamma_{\rm min}=1$.  Dashed line: the contribution of the
  accretion disk luminosity, assumed to a be represented by a
  black--body.  Bottom panel: zoom in the X--ray band.  The solid
  lines corresponds to the modelling with different values of
  $\gamma_{\rm min}$ (as labelled), illustrating that in this source
  the low energy cut--off cannot be significantly larger than unity. }
\label{1127}
\end{figure}

\item {\bf Shell width ---} Another key parameter for the estimate of
  the kinetic powers is $\Delta R'$.  We set $\Delta R^\prime=R$.
  $\Delta R^\prime$ controls $t_{\rm inj}$ and therefore $\gamma_{\rm
    c}$ in the slow cooling regime.  Variability timescales imply that
  $\Delta R^\prime$\ltsima$R$.  Although there is no obvious
    lower limit to $\Delta R^\prime$ which can be inferred from
    observational constraints, the choice of a smaller $\Delta
    R^\prime$ can lead to an incorrect estimate of the observed flux,
    unless the different travel paths of photons originating in
    different parts of the source are properly taken into account.  As
    illustrative case consider a source with $\theta=1/\Gamma$: the
    photons reaching the observer are those leaving the source at
    90$^{o}$ from the jet axis (in the comoving frame).  Assume also
    that the source emits in this frame for a time interval
    $t^\prime_{\rm inj}$.  If $t^\prime_{\rm inj}< R/c$, then a
    (comoving) observer at 90$^{o}$ can detect photons only from a
  `slice' of the source at any given time.  Only when $t^\prime_{\rm
    inj}> R/c$ the entire source can be seen (Chiaberge \& Ghisellini
  1999).  This is the reason to assume $\Delta R^\prime = R$.


\item {\bf Filling factor ---} Our derivations are based on the
  assumption of a single homogeneous emitting region. However it is
  not implausible to imagine that the emitting volume is
  inhomogeneous, with filaments and/or smaller clumps occupying only a
  fraction of the volume.  How would this alter our estimates? As
  illustrative case let us compare the parameters inferred from the
  SED modelling from a region of size $R$ with one filled by $N_{\rm
    c}$ emitting clouds of typical dimension $r$ and density $n_{\rm
    e, c}$.  As the synchrotron and Compton peak frequencies determine
  univocally the value of the magnetic field, in order to model the
  SED the same field have to permeate the clumps.  This in turn fixes
  the same total number of synchrotron emitting leptons.  If the high
  energy component is due to EC the same spectrum is then produced,
  independently of the filling factor. In the case of a dominant SSC
  emission, instead, it is necessary to also require that the ensemble
  of clouds radiate the same total SSC spectrum, i.e.  that each cloud
  has the same scattering optical depth of the whole homogeneous
  region (i.e. $n_{\rm e, c} r \sim n_{\rm e} R$).

  In both cases (SSC and EC) the kinetic power derived by fitting the
  SED is the same, but in the clumped scenario the required Poynting
  flux can be less (since in this case the same magnetic field
  permeates only the emitting clouds).  This thus strengthens our
  conclusions on the relative importance of $L_{\rm B}$ and $L_{\rm
    p}$ at least in the case of BL Lacs.

\end{itemize}

\section{Results}

The model fitting allowed us to derive the intrinsic physical
parameters of the sources as described in Section 3.  The interesting
quantities thus inferred are reported in Table 1 in the Appendix.
In the Appendix we also report the SEDs of all the
  blazars in our sample and the corresponding spectral models.
Histograms reproducing the distributions of
powers for the populations of FSRQs, BL Lacs and TeV sources are shown
in Fig. \ref{isto}.  
As said these estimates refer to a minimum random
Lorentz factor $\gamma_{\rm min}\simeq 1$ (see below) and $L_{\rm p}$
assumes the presence of one proton per emitting lepton.

Different classes of sources (FSRQs, BL Lacs and TeV detected BL Lacs)
form a sequence with respect to their kinetic powers and Poynting flux
distributions.  Within each class, the spread of the distributions is
similar.  

The robust quantity here is $L_{\rm r}$, directly inferred from
observations and rather model independent as it relies only on
$\Gamma$, providing a lower limit to the total flow power.  $L_{\rm
r}$ ranges between $\sim 10^{43}-10^{47}$ erg s$^{-1}$.  $L_{\rm e}$
and $L_{\rm B}$ reach powers of $\sim 10^{46}$ erg s$^{-1}$, while if a proton
component is present $L_{\rm p}\simeq 10^{42}-10^{48}$ erg s$^{-1}$.
Fig. \ref{isto} also shows the distribution of $L_{\rm e, cold}$,
which corresponds to the rest mass of the emitting leptons, neglecting
their random energy, i.e. $L_{\rm e, cold}=L_{\rm e}/\langle \gamma
\rangle$.

In Fig. \ref{isto_e} the distributions of the energetics corresponding
to the powers shown in Fig. \ref{isto} are reported. These have been
simply computed by considering a power `integrated' over the time
duration of the flare, as measured in the observer frame, 
$\Delta R^\prime/(c\delta)$. 
The energy distributions follow the same trends as the powers.

In order to directly compare the different forms of power with respect to 
the radiated one, in Fig. \ref{lr_vs_all} $L_{\rm p}$, $L_{\rm e}$ and
$L_{\rm B}$ are shown as functions of $L_{\rm r}$.

Finally, Fig. \ref{epsilon} shows the ratios 
$\epsilon_{\rm r}\equiv L_{\rm r}/L_{\rm jet}$, 
$\epsilon_{\rm e}\equiv L_{\rm e}/L_{\rm jet}$ and
$\epsilon_{\rm B}\equiv L_{\rm B}/L_{\rm jet}$ 
as functions of $L_{\rm jet}\equiv L_{\rm p}+L_{\rm e}+L_{\rm B}$. 
In general all three ratios tend to be smaller for increasing
$L_{\rm jet}$, the (anti) correlation being most clear for
$\epsilon_{\rm e}$ (mid panel).  
This is the direct consequence of interpreting the trend observed in blazar 
SEDs in terms of cooling efficiency: in the fast cooling regime 
(powerful sources) low energy leptons (and thus small $\epsilon_{\rm e}$) 
are required at any given time.  
Vice--versa, in the less powerful (TeV) BL Lacs 
$\epsilon_{\rm e}$ is close to unity:
indeed in the slow cooling regime the mean
random Lorentz factor of the emitting leptons approaches (and slightly
exceeds in  several cases) $m_{\rm p}/m_{\rm e}$.  
In the latter
sources assuming one proton per emitting lepton results in $L_{\rm
p}\sim L_{\rm e}$ which is also comparable to $L_{\rm r}$, namely
$\epsilon_{\rm r}$ approaches unity at low $L_{\rm p}$.

In the following we discuss more specifically the results for high
(FSRQs) and low power (BL Lacs) blazars.

\begin{figure}
\hskip -0.5 true cm
\psfig{file=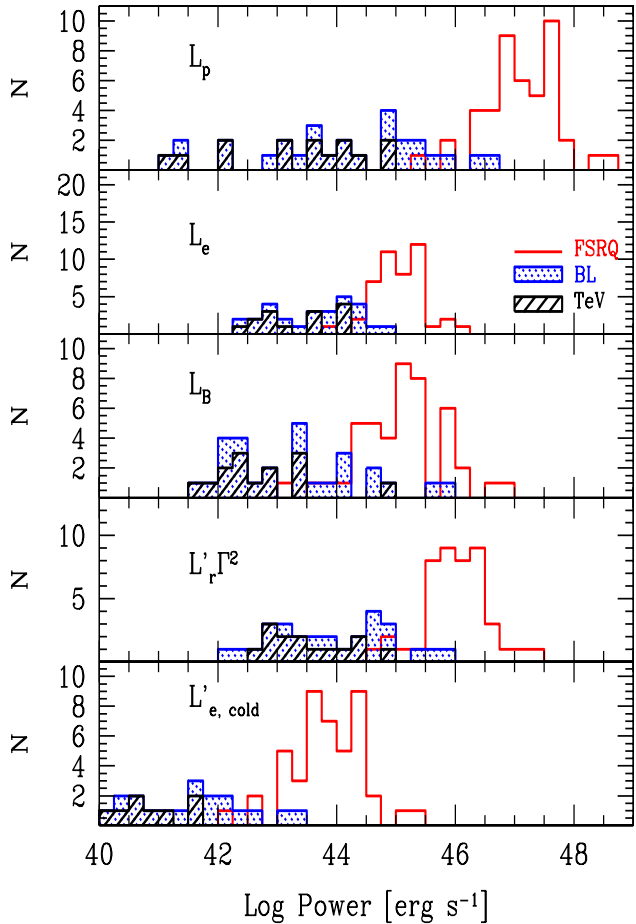,width=9.7truecm,height=13truecm}
\caption{
Power associated to protons, relativistic electrons (or
e$^{\pm}$), Poynting flux, radiatively emitted and bulk energy of cold
leptons.  Hatched areas correspond to BL Lacs and TeV detected
sources.  As detailed in the text these values assume that all leptons
are relativistic, $\gamma_{\rm min} \sim 1$, and that there is one
proton per lepton.}
\label{isto}
\end{figure}
\begin{figure}
\hskip -0.5 true cm
\psfig{file=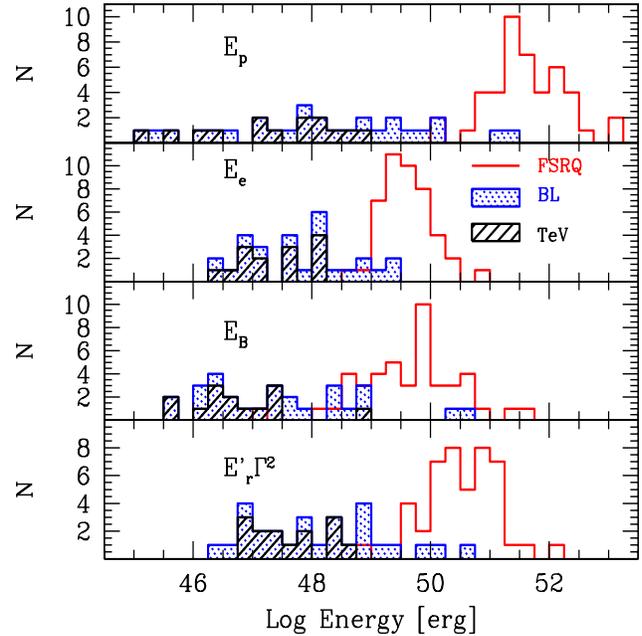,width=9.7truecm,height=11truecm}
\vskip -2 true cm
\caption{
Energy in protons, relativistic leptons, Poynting flux and
emitted radiatively. Hatched areas correspond to BL Lacs.  The
energetics have been calculated multiplying the powers by $\Delta
R^\prime/(\Gamma c)$.}
\label{isto_e}
\end{figure}

\subsection{Flat Spectrum Radio Quasars}

Powerful blazars include FSRQs and some BL Lac objects whose
classification is uncertain, due to the presence of broad (albeit with
small equivalent width) emission lines (e.g. PKS 0537--441).

\begin{figure}
\vskip -0.8 true cm
\hskip -0.5 true cm
\psfig{file=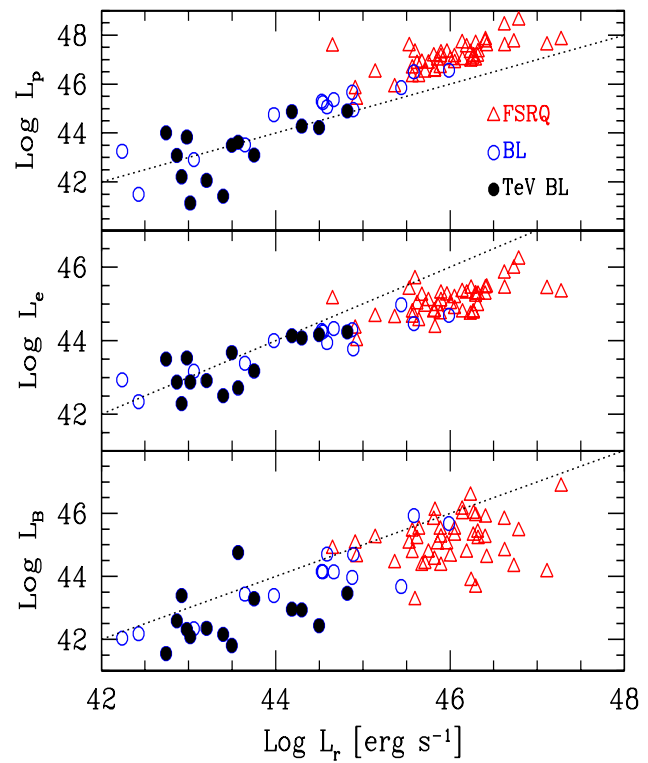,width=9.5truecm,height=12truecm}
\vskip -0.7 true cm
\caption{ Powers associated to the bulk motion of cold protons,
emitting leptons and Poynting flux as functions of the radiative
output $L_{\rm r}$. Triangles: FSRQs, circles: BL Lac objects, filled
circles: TeV detected BL Lacs. The dashed lines correspond to equal
powers.  }
\label{lr_vs_all}
\end{figure}

\begin{figure}
\vskip -0.8 true cm
\hskip -0.5 true cm
\psfig{file=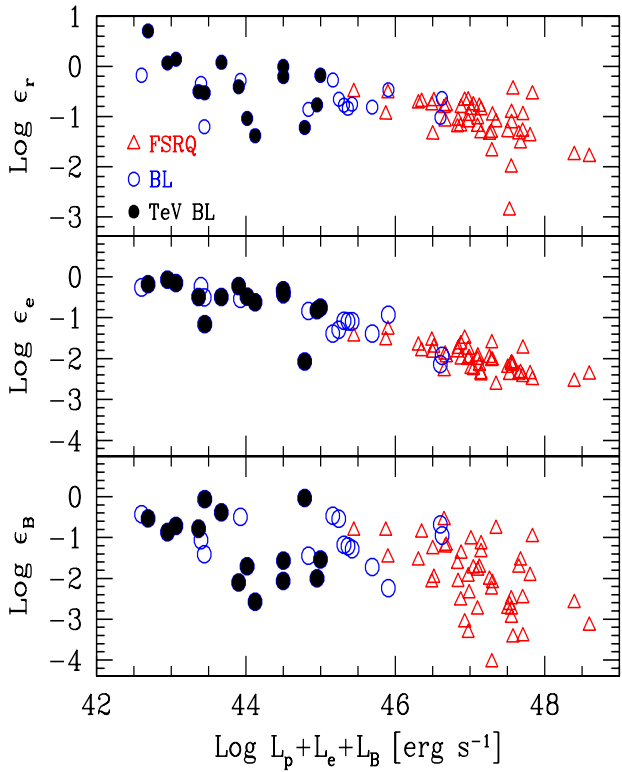,width=9.5truecm,height=12truecm}
\vskip -0.7 true cm
\caption{
The fraction of $L_{\rm jet}$ radiated 
($\epsilon_{\rm r}$, top panel), in relativistic leptons 
($\epsilon_{\rm e}$, mid panel) and in magnetic fields 
($\epsilon_B$, bottom panel) as functions of 
$L_{\rm jet}=L{\rm p}+L_{\rm e}+L_{\rm B}$. 
The TeV BL Lac with efficiency $\epsilon_{\rm r}$
exceeding unity is Mkn 501. 
Symbols as in Fig. \ref{lr_vs_all}.
}
\label{epsilon}
\end{figure}

Fig. \ref{lr_vs_all} shows that the kinetic power
associated to a plasma dominated (in terms of inertia) by relativistic
leptons (electrons and/or e$^{\pm}$) would be typically insufficient
to account for the observed radiation.

As $L_{\rm r}$ exceeds $L_{\rm e}$ and $t_{\rm c}$ for leptons of
all energies is shorter then the dynamical time, the radiating
particles must be continuously injected/re--accelerated.  Thus there
should be another source of power other than that associated to
leptons (see the bottom panel of Fig.~\ref{isto}) able to provide
energy to the emitting particles.

The power in Poynting flux, $L_{\rm B}$, has values comparable to
$L_{\rm e}$ (Fig. \ref{lr_vs_all}). This component
is never dominating, as expected from the fact that the
luminosity of all FSRQs is predominantly in the high energy
component, interpreted as EC emission, which implies that $B$,
controlling the synchrotron output, is limited.  In principle, there
exists a degree of freedom for the estimate of the magnetic field
resulting from the uncertainty on the external radiation energy
density.  The more intense the external radiation density, the larger
the magnetic field, to produce the same Compton to synchrotron
luminosity ratio. Nevertheless $B$ can vary only in a relatively
narrow range, being constrained both by the peak frequency of the
synchrotron component and by the observational limits on the external
photon field if this is due -- as the model postulates -- to broad
line and/or disc photons.

As neither the Poynting flux nor the kinetic power in emitting leptons
are sufficient to account for the radiated luminosity let us then
consider the possible sources of power.

The simplest hypothesis is that jets are loaded with hadrons.  If
there were a proton for each emitting electron, the corresponding
$L_{\rm p}$ would be dominant, a factor $\sim$10--50 larger than
$L_{\rm r}$ (see Fig. \ref{isto} and Fig. \ref{lr_vs_all}).  This
would imply an efficiency $\epsilon_{\rm r}$ ($= L_{\rm r}/L_{\rm p}$)
of order 2--10 per cent.

These efficiencies are what expected if jets supply the radio lobes.
There are two important consequences.  Firstly, a limit on the amount of
electron--positron pairs that can be present.  Since they would lower
the estimated $L_{\rm p}$, only a few (2--3) e$^{\pm}$ per proton are
allowed (see also Sikora \& Madejski 2000).
Secondly, and for the same reason, the lower energy cut--off of
$N(\gamma)$ cannot exceed $\gamma_{\rm min}\sim$ a few.

The inferred values of $L_{\rm p}$ appear to be large if compared to
the {\it average} power required to energise radio lobes (Rawlings \&
Saunders 1991).  However our estimates refer to {\it flaring} states.
To infer average values information on the flare duty cycle would be
needed.  While in general this is not well known, the brightest and
best observed $\gamma$--ray EGRET sources (3C 279, Hartman et
al. 2001, and PKS 0528+134, Ghisellini et al. 1999) indicate values
around 10 per cent (GLAST will provide a excellent estimate on this).
If a duty cycle of 10 per cent is typical of all FSRQs, the average
kinetic powers becomes $\sim$10 times smaller than our estimates, and
comparable with the Eddington luminosity from systems harbouring a
$\sim$ few $10^9$ M$_{\odot}$ black hole.

The power reservoir could be in principle provided also by the inertia
of a population of `cold' (i.e. non emitting) e$^{\pm}$ pairs.  In
order to account for $L_{\rm r}$ -- say to provide $L_{\rm e^{\pm}}
\sim 10^{47}$ erg s$^{-1}$ -- they should amount to a factor
$10^2-10^3$ larger than the number of the radiating particles,
corresponding to a scattering optical depth $\tau_{\rm c} \equiv
\sigma_{\rm T} n_{\rm e^{\pm}} \Delta R' \sim 0.1 L_{47}
\Gamma_1^{-2}\beta^{-1} R_{16}^{-2} \Delta R'_{15}$, where
$\sigma_{\rm T}$ is the Thomson cross section and the value of $R$
refers to the radiating zone\footnote{Throughout this work the notation
$Q=10^xQ_x$ and cgs units are adopted.}.  Conservation of pairs
demands $\tau_{\rm c} \sim 10^3$ at $R \sim 10^{15}$ cm, i.e. the base
of the jet (assuming that there $\Gamma^2\beta\sim 1$).  Such high
values of $\tau_{\rm c}$ however would imply both rapid pair
annihilation (Ghisellini et al. 1992) and efficient interaction with
external photons, leading to Compton drag on the jet and to a visible
spectral component in the X--ray band (Sikora, Begelman \& Rees 1994;
Celotti, Ghisellini \& Fabian 2007).

Within the framework of the assumed model, jets of high power blazars
have then to be heavy, namely dynamically dominated by the bulk motion
of protons, as both leptons and Poynting flux do not provide
sufficient power to account for the observed emission and supply
energy to the radio lobes.  A caveat however is in order, as the
inferred quantities -- in particular the magnetic field intensity --
refer to the emitting region. It is thus not possible to exclude the
presence of a stronger field component whose associated Poynting flux
is energetically dominant.

\subsection{BL Lac objects}

Typically $L_{\rm r}\sim L_{\rm e}$\gtsima$L_{\rm B}$ for BL Lacs.  
This follows the fact that the $\gamma$--ray luminosity in the latter
objects is of the same order (or even larger\footnote{Examples are
1426+428 (Aharonian et al. 2002; Aharonian et al. 2003) and 1101--232
(Aharonian et al. 2006a) once the absorption of TeV photons by the IR
cosmic background is accounted for.})  than the synchrotron one and
for almost all sources the relevant radiation mechanism is SSC,
without a significant contribution from external radiation.  If the
self--Compton process occurred in the Thomson regime then 
$L_{\rm r} \sim L_{\rm B}$, but often the synchrotron seed photons for
the SSC process have high enough energies (UV/X--rays) that the
scattering process is in the Klein--Nishina regime: this implies
$L_{\rm B} < L_{\rm r}$ even for comparable Compton and synchrotron
luminosities.  This result is rather robust indicating that also in BL
Lacs the inferred Poynting luminosity cannot account for the radiated
power on the scales where most of it is produced.

Since $L_{\rm e}\sim L_{\rm r}$, 
relativistic leptons cannot be the
primary energy carriers as they have to be accelerated  in the
radiating zone -- since they would otherwise efficiently cool in the more
compact inner jet region -- at the expenses of another form of energy. 

As before, two the possibilities for the energy reservoir: a cold
leptonic component or hadrons.

The required cold e$^{\pm}$ density is again 10$^2$--$10^3$ times that
in the relativistic population.  Compared to FSRQs, BL Lacs have
smaller jet powers and external photon densities.  Cold e$^{\pm}$
could actually survive annihilation and not suffer significantly of
Compton drag, if the accretion disk is radiatively inefficient.  For
the same reason, these cold pairs would not produce much bulk Compton
radiation (expected in the X--ray band or even at higher energies if
the accretion disk luminosity peaks in the X--rays).

Still the issue of producing these cold pairs in the first place
constitutes a problem.  Electron--photon processes are not efficient
in rarefied plasmas, while photon--photon interactions require a large
compactness at $\sim$MeV energies, where the SED of BL Lacs appears to
have a minimum (although observations in this band do not have high
sensitivity).

Alternatively, also in BL Lacs the bulk energy of hadrons might
constitute the energy reservoir. Even so, one proton per relativistic
lepton provides sometimes barely enough power, since the average
random Lorentz factor of emitting leptons in TeV sources is close to
$m_{\rm p}/m_{\rm e}$ (see Fig. \ref{lr_vs_all}).

This implies either that only a fraction of leptons are accelerated to
relativistic energies (corresponding to $L_{\rm p}$ larger than what
estimated above), or that TeV sources radiatively dissipate most of
the jet power.  If so, their jets have to decelerate. Such option
receives support from VLBI observations showing, in TeV BL Lacs,
sub--luminal proper motion (e.g. Edwards \& Piner 2002; Piner \&
Edwards 2004). And indeed models accounting for the deceleration via
radiative dissipation have been proposed, by e.g.  Georganopoulos \&
Kazanas (2003) and Ghisellini, Tavecchio \& Chiaberge (2005). The
latter authors postulate a spine/layer jet structure that can lead, by
the Compton rocket effect, to effective deceleration even assuming the
presence of a proton per relativistic lepton.  While these models are
more complex than what assumed here it should be stressed that the
physical parameters inferred in their frameworks do not alter the
scenario illustrated here (in these models the derived magnetic field
can be larger, but the corresponding Poynting flux does not dominate
the energetics).

The simplest option is thus that also for low luminosity blazars the
jet power is dominated by the contribution due to the bulk motion of
protons, with the possibility that in these sources a significant
fraction of it is efficiently transferred to leptons and radiated
away.

\begin{figure}
\vskip -0.5true cm
\hskip -2.5 true cm
\psfig{file=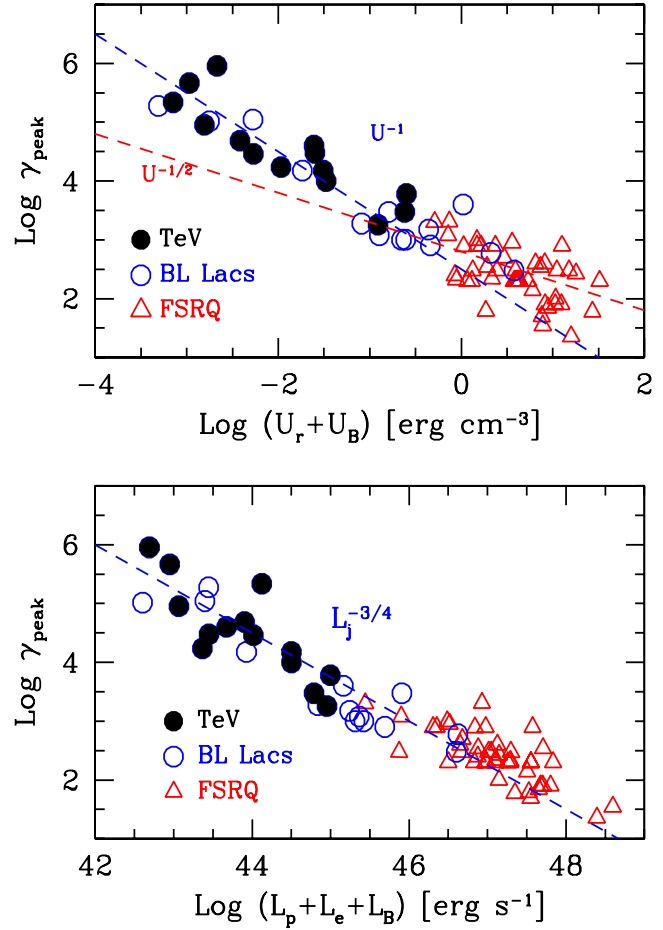,width=13.3truecm,height=14truecm}
\vskip -0.7 true cm
\caption{
Top panel: 
The blazar sequence in the plane $\gamma_{\rm peak}$--$U$ 
($U=U_{\rm r}+U_{\rm B}$).  
The dashed lines corresponding to $\gamma_{\rm peak}\propto U^{-1}$ 
and $\gamma_{\rm peak}\propto U^{-1/2}$ are not formal fits, but 
guides to the eye.
Bottom panel: The blazar sequence in the plane 
$\gamma_{\rm peak}$--$L_{\rm jet}$, where $L_{\rm jet}$ is the sum of the proton,
lepton and magnetic field powers.  
Again, the dashed line $\gamma_{\rm peak} \propto L_{\rm jet}^{-3/4}$ 
is not a formal fit.  Symbols as in Fig. \ref{lr_vs_all}. }
\label{gb}
\end{figure}

\subsection{The blazar sequence}

The dependence of the radiative regime on the source power can be
highlighted by directly considering the random Lorentz factor
$\gamma_{\rm peak}$ of leptons responsible for both peaks of the
emission (synchrotron and inverse Compton components) as a function of
the comoving energy density $U = U_{\rm B}+U_{\rm r}$ (top panel of
Fig. \ref{gb}). $U_{\rm r}$ corresponds to the fraction of the total
radiation energy density available for Compton scattering in the
Thomson regime.  In powerful blazars this coincides with the energy
density of synchrotron and broad line photons, while in TeV BL Lacs it
is a fraction of the synchrotron radiation.

The figure illustrates one of the key features of the blazar sequence,
offering an explanation of the phenomenological trend between the
observed bolometric luminosity and the SED of blazars, as presented in
Fossati et al. (1998) and discussed in G98 and G02.  The inclusion
here of TeV BL Lacs confirms and extends the $\gamma_{\rm peak}$--$U$
relation towards high $\gamma_{\rm peak}$ (low $U$).  The sequence
appears to comprise two branches: the high $\gamma_{\rm peak}$ branch
can be described as $\gamma_{\rm peak}\propto U^{-1}$, while below
$\gamma_{\rm peak} \sim 10^3$ the relation seems more scattered, with
objects still following the above trend and others following a flatter
one, $\gamma_{\rm peak}\propto U^{-1/2}$.

The steep branch can be interpreted in terms of radiative cooling:
when $\gamma_{\rm c}>\gamma_{\rm inj}$, the particle distribution
presents two breaks: 
below $\gamma_{\rm inj}$  $N(\gamma)\propto \gamma^{-1}$, 
between $\gamma_{\rm inj}$ and $\gamma_{\rm c}$ 
$N(\gamma)\propto \gamma^{-(n-1)}$    
(which is the slope of the injected distribution $s=n-1$),
and above $\gamma_{\rm c}$  $N(\gamma)\propto \gamma^{-n}$.
Consequently, for $n<4$, the resulting synchrotron and inverse Compton
spectral peaks are radiated by leptons with 
$\gamma_{\rm peak}=\gamma_{\rm c}$ given by:
\begin{equation}
\gamma_{\rm c} \, =\, {3 \over 4\sigma_{\rm T} U \Delta R^\prime},
\end{equation}
thus accounting for the steeper correlation. The scatter around
the correlation is due to different values of $\Delta R^\prime$ and 
to sources requiring $n>4$, for which $\gamma_{\rm peak}=\gamma_{\rm inj}$ 
(see Table 1).

When $\gamma_{\rm c}<\gamma_{\rm inj}$, instead, all of the injected
leptons cool in the time $t_{\rm inj}=\Delta R^\prime/c$. 
If $n<4$, $\gamma_{\rm peak}$ coincides with $\gamma_{\rm inj}$,
while it is still equal to $\gamma_{\rm c}$ when $n>4$.
This explains why part of the sources still follow
the $\gamma_{\rm peak}\propto U^{-1}$ relation also for
small values of $\gamma_{\rm peak}$.

The physical interpretation of the $\gamma_{\rm peak}\propto U^{-1/2}$
branch is instead more complex, since in this case $\gamma_{\rm
peak}=\gamma_{\rm inj}$, which is a free parameter of the model.  As
discussed in G02, one possibility is that $\gamma_{\rm inj}$
corresponds to a pre--injection phase (as envisaged for internal
shocks in Gamma--Ray Bursts).  During such phase leptons would be
heated up to energies at which heating and radiative cooling balance.
If the acceleration mechanism is independent of $U$ and $\gamma$, the
equilibrium is reached at Lorentz factors $\gamma \propto U^{-1/2}$,
giving raise to the flatter branch.

The trend of a stronger radiative cooling reducing the value of
$\gamma_{\rm peak}$ in more powerful jets is confirmed by considering
the direct dependence of $\gamma_{\rm peak}$ on the total jet power
$L_{\rm jet} = L_{\rm p}+ L_{\rm e}+ L_{\rm B}$.  This is reported in
the bottom panel of Fig. \ref{gb}.   The correlation approximately
  follows the trend $\gamma_{\rm peak}\propto L_{\rm jet}^{-3/4}$ and
  has a scatter comparable to that of the $\gamma_{\rm peak}$--$U$
  relation.  

\subsection{The outflowing mass rate}

The inferred jet powers and the above considerations supporting the
dominant role of $L_{\rm p}$ allow to estimate a mass outflow rate,
$\dot M_{\rm out}$, corresponding to flaring states of the
sources, from
\begin{equation}
L_{\rm p}\, =\, \dot M_{\rm out} \Gamma c^2\, \to \, \dot M_{\rm out}
\, = \, {L_{\rm p}\over \Gamma c^2} \, \simeq 0.2 {L_{\rm p, 47} \over
\Gamma_1}\,\, {M_\odot \over \rm yr}.
\end{equation}

A key physical parameter is given by the ratio between $\dot M_{\rm out}$ and
the mass accretion rate, $\dot M_{\rm in}$, that can be derived by the
accretion disk luminosity: $L_{\rm disk}=\eta \dot M_{\rm in} c^2$,
where $\eta$ is the radiative efficiency:
\begin{equation}
{\dot M_{\rm out} \over \dot M_{\rm in} }  \, =\, 
{\eta\over \Gamma} \, {L_{\rm p} \over L_{\rm disk}} \, = \, 
10^{-2} \, {\eta_{-1} \over \Gamma_1}\, {L_{\rm p} \over L_{\rm disk}}.
\end{equation}

Rawlings \& Saunders (1991) argued that the average jet power required
to energise radio lobes is of the same order of the accretion disk
luminosity as estimated from the narrow lines emitted following 
photoionization (see also
Celotti, Padovani, \& Ghisellini 1997 who considered broad lines to
infer the disc emission). Here jet powers in general larger than
the accretion disk luminosity have been instead inferred: for powerful
blazars with broad emission lines the estimated ratio $L_{\rm
p}/L_{\rm disk}$ is of order 10--100 (see Table 2).  As in
these systems typically $\Gamma\sim 15$
and for accretion efficiencies $\eta\sim0.1$, inflow and outflow
mass rates appear to be comparable during flares.

A challenge for the $\gamma$--ray satellite GLAST will be to reveal
whether low--quiescent states of activity correspond to episodes of
lower radiative efficiency or reduced $L_{\rm p}$ and in the latter
case to distinguish if a lower $L_{\rm p}$ is predominantly determined
by a lower $\dot M_{\rm out}$ or $\Gamma$.

\subsection{Summary of results}

\begin{itemize}
\item
The estimated jet powers often exceed the power radiated by accretion,
which can be derived directly for the most powerful sources, whose
synchrotron spectrum peaks in the far IR, and via the luminosity of
the broad emission lines in less powerful FSRQs (see e.g. Celotti,
Padovani \& Ghisellini 1997; Maraschi \& Tavecchio 2003).
\item 
For powerful blazars (i.e. FSRQs) the radiated luminosity is
in some cases larger than the power carried in the relativistic leptons
responsible for the emission.
\item
Also the values of the Poynting flux are statistically lower 
than the radiated power.
This directly follows from the dominance of the Compton over the
synchrotron emission.
\item
If there is a proton for each emitting electron, the kinetic power
associated to the bulk motion in FSRQs is a factor 10--50 larger than
the radiated one, i.e. corresponding to efficiencies of 2--10 per
cent. This is consistent with a significant fraction being able to
energise radio lobes.  The proton component has to be energetically
dominant (only a few electron--positron pairs per proton are allowed)
unless the magnetic field present in the emitting region is only a
fraction of the Poynting flux associated to jets.
\item 
For low power BL Lacs the power in relativistic leptons is
comparable to the emitted one.  Nevertheless, an additional reservoir
of energy is needed to accelerate them to high energies.  This
cannot be the Poynting flux, that again appears to be insufficient.
\item
The contribution from kinetic energy of protons is an obvious
candidate, but since the average random Lorentz factors of leptons can
be as high as 
$\langle \gamma\rangle \sim 2000\sim m_{\rm p}/m_{\rm e}$ in TeV sources, 
one proton per emitting electrons yields $L_{\rm p}\sim L_{\rm e}$.
\item
This suggests that either only a fraction of leptons are accelerated
to relativistic energies or that jets dissipate most of their bulk
power into radiation.  In the latter case they should decelerate.
\item
The jet power (inversely) correlates with the energy of the leptons
emitting at the peak frequencies of the blazar SEDs.  This indicates
that radiative cooling is most effective in more powerful jets.
\item
The need for a dynamically dominant proton component in blazars allows
to estimate the mass outflow rate $\dot M_{\rm out}$.  This reaches,
during flares, values comparable to the mass accretion rate.
\end{itemize}

\section{Discussion}

The first important result emerging from this work is that the power
of extragalactic jets is large in comparison to that emitted via
accretion.  This result is rather robust, since the uncertainties
related to the particular model adopted are not crucial: the finding
follows from a comparison with the emitted luminosity, which is a
rather model--independent quantity, relying only on the estimate of
the bulk Lorentz factor.  The findings about the kinetic and
  Poynting powers instead depend on the specific modelling of the
  blazar SEDs as synchrotron and Inverse Compton emission from a
  one--zone homogeneous region.  Hadronic models may yield different
  results.  Furthermore, the estimated power associated to the proton
  bulk motion relies also on the amount of `cold' (non emitting)
  electron--positron pairs in the jet.  We have argued that if pairs
  had to be dynamically relevant their density at the jet base would
  make annihilation unavoidable. However the presence of a few pairs per
  proton cannot be excluded.  If there were no electron--positron
  pairs, the inferred jet powers are 10--100 times larger than the
disk accretion luminosity, in agreement with earlier claims based on
individual sources or smaller blazar samples (Ghisellini 1999; Celotti
2001; Maraschi \& Tavecchio 2003; Sambruna et al. 2006).  Such large
powers are needed in order to energise the emitting leptons at the
($\gamma$--ray) jet scale and the radio--lobes hundreds of kpc away.


The finding that blazar jets are not magnetically dominated is also
quite robust, but only in the context of the (widely accepted)
framework of the synchrotron--inverse Compton emission model.  In this
scenario the dominance of the high energy (inverse Compton) component
with respect to the synchrotron one limits the magnetic field.  This
is at odd with magnetically driven jet acceleration, though this
appears to be the most viable possibility. In blazars thermally driven
acceleration, as invoked in Gamma--Ray Bursts, does not appear to be
possible. In Gamma-Ray Bursts the initial fireball is highly opaque to
electron scattering and this allows the conversion of the trapped
radiation energy into bulk motion (see e.g. Meszaros 2006 for a recent
review).  In blazars the scattering optical depths at the base of the
jet are around unity at most, and even invoking the presence of
electron--positron pairs to increase the opacity is limited by the
fact that they quickly annihilate.  Thus, if magnetic fields play a
crucial role our results would require that magnetic acceleration must
be rapid, since at the scale of a few hundreds Schwarzschild radii,
where most of radiation is produced, the Poynting flux is no longer
energetically dominant (confirming the results by Sikora et al. 2005).
However models of magnetically accelerated flows indicate that the
process is actually relatively slow (e.g. Li, Chiueh \& Begelman 1992,
Begelman \& Li 1994). Apparently the only possibility is that the jet
structure is more complex than what assumed and a possibly large
scale, stronger field does not pervade the dissipation region, as also
postulated in pure electromagnetic scenarios (see e.g. Blandford 2002,
Lyutikov \& Blandford 2003).

The third relevant results refers the difference between FSRQs and BL
Lacs. This concerns not only their jet powers but also the relative
role of protons in their jets. BL Lacs would be more dissipative and
therefore their jets should decelerate.  This inference depends on
assuming one proton per emitting lepton also in these sources, and
this is rather uncertain (i.e. there could be more than one proton per
relativistic, emitting electron).  If true, it can provide an
explanation to why VLBI knots of low power BL Lacs are moving
sub-luminally and in turn account for the different radio morphology
of FR I and FR II radio--galaxies, since low power BL Lacs are
associated to FR I sources.

\section*{Acknowledgements}  We thank the referee, Marek Sikora,
  for his very constructive comments and Fabrizio Tavecchio for
  discussions. The Italian MIUR is acknowledged for financial support
  (AC).

\begin{table*}
\caption{The input parameters of the model for FSRQs.
(1) Source name; 
(2) redshift;
(3) radius $R$ of emitting region in units of $10^{15}$ cm; 
(4) intrinsic injected power in units of $10^{45}$ erg s$^{-1}$;
(5) bulk Lorentz factor;
(6) viewing angle;
(7) magnetic field intensity (in Gauss); 
(8) minimum random Lorentz factor of the injected particles; 
(9) maximum random Lorentz factor of the injected particles; 
(10) $\gamma_{\rm peak}$; 
(11) spectral slope of particles above the cooling break; 
(12) disk luminosity in units of $10^{45}$ erg s$^{-1}$;
(13) radius of the BLR in units of $10^{15}$ cm.
(14) random Lorentz factor of the electrons cooling in $\Delta R^\prime/c$.
}
\begin{tabular}{@{}llccccccccccccl}
\hline 
Source &$z$ &$R$ &$L^\prime_{\rm inj}$ &$\Gamma$ &$\theta$ &$B$ &$\gamma_{\rm inj}$ 
&$\gamma_{\rm max}$ &$\gamma_{\rm peak}$ &$n$ &$L_{\rm d}$ &$R_{\rm BLR}$ &$\gamma_{\rm c}$ \\ 
(1) &(2) &(3) &(4) &(5) &(6) &(7) &(8) &(9) &(10) &(11) &(12) &(13) &(14) \\ 
\hline
0202$+$149       &0.405 &5  &5.0e-2 &17 &5.0 &0.8  &100  &1.8e3 &210   &3.2 &4.e-2 &250 &210   \\ 
0208$-$512       &1.003 &8  &3.0e-2 &15 &2.6 &2.5  &900  &4.0e4 &900   &4.0 &2     &240 &32    \\ 
0234$-$285       &1.213 &20 &2.5e-2 &16 &3.0 &4.0  &100  &3.0e3 &100   &3.78&5     &220 &4.3   \\ 
0336$-$019       &0.852 &20 &1.0e-1 &16 &3.0 &1.0  &200  &4.0e3 &200   &3.7 &7     &400 &10.7  \\ 
0420$-$014       &0.915 &20 &3.0e-2 &16 &3.0 &2.8  &500  &6.0e3 &500   &3.5 &1.5   &240 &15.8  \\ 
0440$-$003       &0.844 &20 &1.8e-2 &16 &3.3 &2.8  &800  &1.0e4 &800   &3.5 &1     &280 &28.1  \\ 
0446$+$112       &1.207 &15 &8.0e-2 &16 &3.0 &0.45 &800  &1.0e4 &800   &3.7 &1.5   &250 &26.2  \\ 
0454$-$463       &0.858 &15 &1.5e-2 &16 &3.5 &1.6  &800  &1.0e4 &800   &3.7 &1     &270 &41.5  \\ 
0521$-$365       &0.055 &8  &1.0e-2 &15 &9.0 &2.2  &1.2e3&1.5e4 &1.2e3 &3.2 &2.e-2 &200 &161   \\ 
0528$+$134       &2.07  &30 &7.5e-1 &16 &3.5 &9.0  &200  &1.0e4 &200   &3.5 &40    &370 &1     \\ 
0804$+$499       &1.433 &20 &7.0e-2 &16 &3.5 &5.0  &270  &3.0e3 &270   &4.1 &20    &340 &2.6   \\ 
0805$-$077       &1.837 &15 &8.0e-2 &15 &3.0 &3.2  &300  &3.0e3 &300   &3.4 &27    &400 &4.1   \\ 
0827$+$234       &2.05  &15 &6.0e-2 &15 &3.0 &7.0  &300  &3.0e3 &300   &3.4 &16    &400 &5.6   \\ 
0836$+$710       &2.172 &18 &0.4    &14 &2.6 &3.4  &35   &6.0e3 &35    &3.7 &20    &500 &6.6   \\ 
0906$+$693       &5.47  &15 &0.4    &18 &2.5 &0.7  &800  &6.0e3 &800   &3.7 &50    &700 &4.9   \\ %
0917$+$449       &2.18  &10 &5.0e-2 &15 &3.0 &6.0  &350  &3.0e3 &350   &3.1 &9     &400 &12.6  \\ 
0954$+$556       &0.901 &20 &5.0e-3 &15 &3.0 &1.1  &2.0e3&8.0e5 &2.0e3 &3.7 &0.5   &300 &90.8  \\ 
1127$-$145       &1.187 &25 &6.0e-2 &18 &2.5 &3.3  &70   &2.0e3 &70    &3.4 &12    &420 &4.2   \\ 
1156$+$295       &0.729 &24 &3.0e-2 &15 &2.7 &5.0  &400  &6.0e3 &400   &3.4 &10    &400 &6.0   \\ 
1222$+$216       &0.435 &20 &6.0e-3 &15 &4.0 &2.2  &200  &6.0e3 &200   &3.9 &1     &300 &39.7  \\ 
1226$+$023       &0.158 &6  &6.0e-2 &12 &5.0 &7.5  &50   &6.0e3 &50    &4.2 &25    &600 &20.4  \\ 
1229$-$021       &1.045 &10 &4.0e-2 &15 &4.0 &4.5  &200  &6.0e3 &200   &4.4 &8     &500 &21.3  \\ 
1253$-$055       &0.538 &22 &5.0e-2 &15 &3.5 &2.2  &250  &2.0e3 &250   &3.2 &3.5   &400 &19.4  \\ 
1313$-$333       &1.210 &20 &2.5e-2 &15 &3.5 &1.3  &200  &3.0e3 &775   &3.0 &1     &300 &43.5  \\ 
1334$-$127       &0.539 &15 &5.5e-3 &12 &4.0 &3.0  &300  &4.0e3 &300   &3.9 &2     &350 &46.4  \\ 
1406$-$076       &1.494 &17 &8.0e-2 &15 &3.3 &0.54 &700  &6.0e3 &2.5e3 &3.0 &0.7   &300 &73.4  \\ 
1424$-$418       &1.522 &18 &8.0e-2 &16 &3.3 &2.8  &400  &4.0e3 &400   &3.8 &20    &500 &6.3   \\ 
1510$-$089       &0.361 &8  &2.0e-3 &16 &2.7 &3.5  &10   &2.0e3 &62.4  &3.7 &1.3   &310 &62.4  \\ 
1606$+$106       &1.226 &15 &3.0e-2 &16 &2.7 &1.0  &200  &2.0e3 &200   &3.7 &10    &500 &15.7  \\ 
1611$+$343       &1.404 &15 &2.8e-2 &16 &2.7 &2.2  &200  &2.0e3 &200   &3.3 &10    &500 &14.9  \\ 
1622$-$253       &0.786 &15 &1.9e-2 &16 &4.0 &1.0  &250  &3.0e3 &250   &3.4 &0.7   &300 &72.8  \\ 
1622$-$297       &0.815 &13 &7.0e-1 &16 &4.0 &1.1  &350  &2.5e3 &350   &3.1 &0.7   &300 &37.2  \\ 
1633$+$382       &1.814 &20 &1.5e-1 &17 &2.6 &1.2  &200  &7.0e3 &200   &3.2 &12    &500 &8.6   \\ 
1730$-$130       &0.902 &20 &1.6e-2 &16 &3.0 &2.0  &200  &4.0e3 &200   &3.4 &4     &600 &35.5  \\ 
1739$+$522       &1.375 &15 &4.0e-2 &16 &3.0 &1.4  &200  &5.0e3 &200   &3.1 &6     &400 &16.4  \\ 
1741$-$038       &1.054 &15 &3.5e-2 &16 &3.0 &2.2  &200  &5.0e3 &200   &4.6 &8     &450 &15.0  \\ 
1830$-$211       &2.507 &20 &6.5e-2 &15 &3.0 &1.3  &140  &4.0e3 &140   &4.1 &7     &320 &7.7   \\ 
1933$-$400       &0.965 &15 &1.4e-2 &16 &3.7 &3.5  &300  &3.0e3 &300   &3.6 &3     &400 &25.2  \\ 
2052$-$474       &1.489 &20 &8.0e-2 &16 &3.7 &2.0  &300  &3.0e3 &300   &3.6 &6     &400 &11.9  \\ 
2230$+$114       &1.037 &20 &5.0e-2 &17 &3.0 &5.5  &80   &1.0e4 &80    &4.0 &10    &400 &5.6   \\ 
2251$+$158       &0.859 &30 &7.0e-2 &16 &3.5 &6.5  &60   &4.0e4 &60    &3.4 &30    &340 &1.1   \\ 
2255$-$282       &0.926 &10 &2.0e-2 &16 &2.8 &1.6  &1000 &2.5e3 &1000  &3.7 &2     &400 &62.5  \\ 
\hline
0525$-$3343      &4.41  &26 &8.0e-2 &17 &2.8 &1.5  &80   &2.0e3 &80    &3.7 &130   &1100 &2.9  \\ 
1428$+$4217      &4.72  &20 &1.5e-1 &16 &3.2 &4.0  &23   &2.0e3 &23    &3.5 &70    &700  &2.9  \\ 
1508$+$5714      &4.3   &20 &1.3e-1 &14 &3.5 &5.0  &80   &4.0e3 &80    &3.7 &150   &1100 &4.4  \\ 
\hline
\hline
\end{tabular}
\end{table*}

\setcounter{table}{0}
\begin{table*}
\caption{
(continue) --
The input parameters of our model for BL Lac objects. Columns 1--14 as in 
the first part of the Table.
Column (15): LBL=Low energy peak BL Lacs, HBL=High energy peak BL Lac,
TeV=BL Lacs detected in the TeV band (all are also HBLs).
} 
\begin{tabular}{@{}llccccccccccccl}
\hline 
Source &$z$ &$R$ &$L^\prime_{\rm inj}$ &$\Gamma$ &$\theta$ &$B$ &$\gamma_{\rm inj}$ 
&$\gamma_{\rm max}$ &$\gamma_{\rm peak}$ &$n$ &$L_{\rm d}$ &$R_{\rm BLR}$ &$\gamma_{\rm c}$ &Class\\
(1) &(2) &(3) &(4) &(5) &(6) &(7) &(8) &(9) &(10) &(11) &(12) &(13) &(14) &(15)\\ 
\hline
0033$+$595      &0.086 &5  &1.5e-5 &20 &1.8 &0.2  &7.0e4 &1.0e6 &1.0e5 &3.1  &---   &---   &1.0e5 &HBL \\ 
0120$+$340      &0.272 &4  &3.6e-5 &24 &1.5 &0.35 &1.0e4 &2.8e5 &1.1e5 &3.0  &---   &---   &4.4e5 &HBL \\  
0219$+$428      &0.444 &6  &3.0e-3 &16 &3.0 &3.8  &4.0e3 &6.0e4 &4.0e3 &3.6  &---   &---   &147   &LBL\\
0235$+$164      &0.940 &25 &5.0e-2 &15 &3.0 &3.0  &600   &2.5e4 &600   &3.2  &3     &400   &17.6  &LBL \\
0537$-$441      &0.896 &20 &1.7e-2 &15 &3.0 &5.0  &300   &5.0e3 &300   &3.4  &5     &400   &12.3  &LBL  \\
0548$-$322      &0.069 &10 &1.0e-5 &17 &2.4 &0.1  &2.0e3 &1.0e6 &1.9e5 &3.3  &---   &---   &1.9e5 &HBL \\  
0716$+$714   &$>$0.3   &8  &1.3e-3 &17 &2.6 &2.7  &1.5e3 &2.5e4 &1.5e3 &3.4  &---   &---   &264   &LBL \\
0735$+$178   &$>$0.424 &8  &2.0e-3 &15 &2.6 &1.6  &1.0e3 &9.0e3 &1.0e3 &3.2  &---   &---   &468   &LBL \\
0851$+$202     &0.306  &8  &1.5e-3 &15 &2.6 &1.6  &1.0e3 &9.0e3 &1.0e3 &3.2  &---   &---   &521   &LBL \\
0954$+$658     &0.368  &15 &2.0e-3 &13 &3.5 &1.0  &1.2e3 &5.0e3 &1.2e3 &3.5  &0.08  &300   &484   &LBL \\
1114$+$203     &0.139  &10 &1.5e-4 &17 &2.5 &0.5  &1.5e4 &3.0e5 &1.5e4 &4.5  &---   &---   &5.0e3 &HBL  \\
1219$+$285     &0.102  &6  &4.0e-4 &15 &3.3 &0.9  &1.5e3 &6.0e4 &1.9e3 &3.8  &---   &---   &1.9e3 &LBL \\
1604$+$159     &0.357  &15 &3.7e-3 &15 &3.8 &0.7  &800   &5.0e4 &800   &3.6  &0.2   &200   &134   &LBL  \\ 
2032$+$107     &0.601  &10 &1.0e-2 &16 &3.6 &0.7  &3.0e3 &1.0e5 &3.0e3 &4.3  &---   &---   &576   &LBL  \\ 
\hline
1011$+$496     &0.212  &6  &1.2e-3 &20 &1.7 &0.3  &1.0e4 &4.0e5 &1.2e4 &4.2  &---   &---   &1.2e4 &TeV \\   
1101$-$232     &0.186  &6  &2.0e-4 &20 &1.7 &0.15 &4.0e4 &1.5e6 &4.7e5 &3.0  &---   &---   &1.4e5 &TeV \\  
1101$+$384     &0.031  &6  &4.0e-5 &18 &2.0 &0.09 &1.0e3 &4.0e5 &2.2e5 &3.2  &---   &---   &2.2e5 &TeV \\  
1133$+$704     &0.046  &6  &3.5e-5 &17 &3.5 &0.23 &4.0e3 &8.0e5 &2.9e4 &3.9  &---   &---   &2.9e4 &TeV \\    
1218$+$304     &0.182  &6  &2.0e-4 &20 &2.7 &0.6  &4.0e4 &7.0e5 &4.0e4 &4.0  &---   &---   &6.3e3 &TeV \\    
1426$+$428     &0.129  &5  &2.0e-4 &20 &2.2 &0.13 &1.0e4 &5.0e6 &4.8e4 &3.3  &---   &---   &4.8e4 &TeV \\  
1553$+$113  &$>$0.36   &4  &1.6e-3 &20 &1.8 &1.1  &6.0e3 &4.0e5 &6.0e3 &4.0  &---   &---   &920   &TeV \\     
1652$+$398     &0.0336 &7  &1.2e-3 &14 &3.0 &0.2  &9.0e5 &4.0e6 &9.0e5 &3.2  &---   &---   &6.2e4 &TeV \\  
1959$+$650     &0.048  &6  &2.9e-5 &18 &2.5 &0.75 &3.0e4 &3.0e5 &3.0e4 &3.1  &---   &---   &6.1e3 &TeV \\     
2005$-$489     &0.071  &9  &1.1e-4 &18 &2.6 &2.4  &3.0e3 &8.0e5 &3.0e3 &3.3  &---   &---   &427 &TeV \\  
2155$-$304     &0.116  &5  &9.0e-4 &20 &1.7 &0.27 &1.5e4 &2.0e5 &1.5e4 &3.5  &---   &---   &5.9e3 &TeV \\ 
2200$+$420     &0.069  &5  &8.0e-4 &14 &3.3 &0.7  &1.8e3 &1.0e6 &1.8e3 &3.9  &2.5e-2&200   &1.5e3 &TeV \\ 
2344$+$512     &0.044  &5  &4.0e-5 &16 &4.0 &0.4  &3.0e3 &9.0e5 &1.7e4 &3.1  &---   &---   &1.7e4 &TeV \\ 
2356$-$309     &0.165  &8  &2.5e-4 &18 &2.6 &0.17 &9.0e4 &3.0e6 &9.0e4 &3.1  &---   &---   &7.4e4 &TeV \\  

\hline
\end{tabular}
\end{table*}

\begin{table*}
\caption{Kinetic powers and Poynting fluxes 
(all in units of $10^{45}$ erg s$^{-1}$) ---
(1) Source name; 
(2) Total (synchrotron + IC) radiative power $L_{\rm r}$;
(3) Synchrotron radiative power $L_{\rm s}$;
(4) Poynting flux $L_{\rm B}$;
(5) Kinetic power in emitting electrons $L_{\rm e}$;
(6) Kinetic power in protons $L_{\rm p}$, assuming one proton per electron;
(7) Average random electron Lorentz factor $\langle\gamma\rangle$
}
\begin{tabular}{@{}lllllll}
\hline 
Source &$L_{\rm r}$   &$L_{\rm s}$  &$L_{\rm B}$  &$L_{\rm e}$  &$L_{\rm p}$  &$\langle\gamma\rangle$  \\
(1) &(2) &(3) &(4) &(5) &(6) &(7) \\ 
\hline
0202$+$149    &3.96    &8.5e-2  &1.7e-2 &4.66   &191.5    &44.6  \\    
0208$-$512    &6.63    &0.45    &0.337  &0.61   &31.92    &34.9  \\
0234$+$285    &6.49    &0.23    &6.132  &0.58   &133.4    &8.0   \\
0336$-$019    &26.2    &0.16    &0.383  &2.76   &355.2    &14.3  \\
0420$-$014    &7.71    &0.78    &3.00   &0.52   &44.5     &21.4  \\
0440$-$003    &4.29    &0.87    &3.00   &0.34   &19.2     &32.5  \\
0446$+$112    &19.6    &6.5e-2  &4.4e-2 &1.56   &92.9     &30.8  \\   
0454$-$463    &3.69    &0.26    &0.55   &0.42   &19.4     &40.2  \\
0521$-$365    &3.32    &0.63    &0.26   &0.41   &7.32     &102   \\
0528$+$134    &188.3   &16.1    &69.8   &2.04   &612      &6.1   \\ 
0804$+$499    &18.6    &0.83    &9.58   &0.55   &130.6    &7.7   \\
0805$-$077    &18.5    &0.44    &1.94   &0.58   &108.5    &9.9   \\   
0827$+$243    &13.9    &2.22    &9.28   &0.58   &93.3     &11.4  \\ 
0836$+$710    &61.3    &0.96    &2.75   &16.0   &3940     &7.5   \\
0906$+$693    &129.1   &0.19    &0.13   &2.50   &373      &12.3  \\
0917$+$449    &11.3    &2.03    &3.03   &0.72   &72.1     &18.3  \\
0954$+$556    &0.85    &0.10    &0.41   &9.64e-2&2.27     &77.9  \\   
1127$-$145    &19.7    &0.55    &8.26   &1.81   &440.5    &7.5   \\
1156$+$295    &6.70    &1.00    &12.12  &0.22   &32.7     &12.4  \\
1222$+$216    &1.38    &0.18    &1.63   &0.44   &29.7     &27.3  \\
1226$+$023    &3.39    &0.28    &1.09   &2.43   &353.0    &12.6  \\
1229$-$021    &7.84    &0.94    &1.70   &1.90   &179.0    &19.4  \\
1253$-$055    &11.25   &0.84    &1.97   &1.43   &123.8    &21.2  \\
1313$-$333    &5.63    &0.30    &0.57   &1.19   &67.6     &32.2  \\
1334$-$127    &0.81    &0.19    &1.09   &0.21   &6.18     &62.6  \\
1406$-$076    &17.4    &0.27    &7.1e-2 &2.60   &82.5     &58.0  \\
1424$-$418    &20.7    &0.68    &2.43   &0.88   &130.2    &12.4  \\  
1510$-$089    &0.45    &4.6e-2  &0.75   &1.34   &335.1    &7.3   \\
1606$+$106    &7.84    &5.3e-2  &0.22   &1.17   &124.9    &17.3  \\
1611$+$343    &7.19    &0.26    &1.04   &0.88   &92.4     &17.4  \\
1622$-$253    &4.74    &0.18    &0.21   &1.66   &73.1     &41.6  \\  
1622$-$297    &53.9    &1.16    &0.20   &9.02   &502.4    &33.0  \\
1633$+$382    &42.2    &0.38    &0.62   &2.60   &355.5    &13.4  \\
1730$-$130    &4.18    &0.42    &1.53   &0.97   &65.5     &27.2  \\
1739$+$522    &10.1    &0.18    &0.42   &1.00   &96.0     &19.0  \\
1741$-$038    &9.17    &0.26    &1.04   &169    &191.8    &16.2  \\
1830$-$211    &15.4    &8.8e-2  &0.57   &1.90   &317.8    &11.0  \\
1933$-$400    &3.68    &0.64    &2.64   &0.58   &43.2     &24.6  \\
2052$-$474    &21.0    &0.69    &1.53   &1.76   &196.1    &16.5  \\
2230$+$114    &13.7    &0.98    &13.1   &2.11   &459.5    &8.4   \\
2251$+$158    &17.1    &0.76    &36.43  &0.50   &186.2    &5.0   \\
2255$-$282    &5.15    &0.40    &0.25   &0.84   &29.7     &52.0  \\
\hline
0525$-$3343   &25.2    &8.5e-2  &1.65   &1.79   &504.4    &6.5   \\   
1428$+$4217   &41.9    &0.50    &6.13   &6.71   &2464     &5.0   \\ 
1508$+$5714   &25.5    &1.22    &7.33   &2.64   &627.6    &7.7   \\
\hline
\end{tabular}
\end{table*}

\setcounter{table}{1}
\begin{table*}
\caption{
(continue) ---
(1) Source name; 
(2) Total (synchrotron + IC) radiative power $L_{\rm r}$;
(3) Synchrotron radiative power $L_{\rm s}$;
(4) Poynting flux $L_{\rm B}$;
(5) Kinetic power in emitting electrons $L_{\rm e}$;
(6) Kinetic power in protons $L_{\rm p}$, assuming one proton per electron;
(7) Average random electron Lorentz factor $\langle\gamma\rangle$
}
\begin{tabular}{@{}lllllll}
\hline 
Source &$L_{\rm r}$   &$L_{\rm s}$  &$L_{\rm B}$  &$L_{\rm e}$  &$L_{\rm p}$  &$\langle\gamma\rangle$  \\
(1) &(2) &(3) &(4) &(5) &(6) &(7) \\ 
\hline
0033$+$595    &2.67e-3 &2.45e-3 &1.50e-3 &2.22e-3  &3.11e-4  &1.3e4 \\   
0120$+$340    &1.15e-2 &5.94e-3 &2.16e-3 &1.48e-2  &8.22e-3  &3299  \\ 
0219$+$428    &0.78    &0.42    &0.50    &5.99e-2  &0.90     &122   \\
0235$+$164    &9.70    &1.89    &4.73    &0.49     &37.0     &24.3  \\
0537$-$441    &3.84    &0.95    &8.41    &0.29     &21.5     &17.0  \\
0548$-$322    &1.72e-3 &1.36e-3 &1.08e-3 &8.64e-3  &1.81e-2  &877.3 \\
0716$+$714    &0.38    &0.26    &0.50    &8.87e-2  &1.16     &140   \\
0735$+$178    &0.46    &0.19    &0.14    &0.21     &2.29     &172   \\
0851$+$202    &0.35    &0.16    &0.14    &0.17     &1.74     &182   \\
0954$+$658    &0.34    &0.11    &0.14    &0.19     &1.99     &175   \\
1114$+$203    &4.4e-2  &2.42e-2 &2.70e-2 &2.45e-2  &3.32e-2  &1355  \\
1219$+$285    &9.5e-2  &3.76e-2 &2.45e-2 &0.10     &0.56     &325   \\   
1604$+$159    &0.75    &3.59e-2 &9.28e-2 &0.20     &4.62     &79    \\
2032$+$107    &2.74    &0.33    &4.69e-2 &0.95     &7.07     &247   \\
\hline
1011$+$496    &4.86e-2 &1.34e-2 &4.85e-3 &5.90e-2  &6.80e-2  &1595  \\     
1101$-$232    &1.04e-2 &9.13e-3 &1.21e-3 &7.58e-3  &1.37e-4  &1.0e5 \\ 
1101$+$384    &5.51e-3 &2.05e-3 &3.54e-4 &3.17e-2  &0.10     &576   \\ 
1133$+$704    &9.55e-3 &3.75e-3 &2.06e-3 &3.39e-2  &6.80e-2  &917   \\     
1218$+$304    &5.64e-2 &3.19e-2 &1.94e-2 &1.50e-2  &1.26e-2  &2189  \\   
1426$+$428    &3.13e-2 &7.10e-3 &6.33e-4 &4.77e-2  &3.17e-2  &2760  \\
1553$+$113    &0.662   &0.13    &2.90e-2 &0.175    &0.793    &404   \\ 
1652$+$398    &2.50e-2 &1.84e-2 &1.43e-3 &3.21e-3  &2.61e-4  &2.2e4 \\ 
1959$+$650    &8.31e-3 &7.58e-3 &2.46e-2 &1.95e-3  &1.63e-3  &2190  \\
2005$-$489    &3.70e-2 &3.54e-2 &0.57    &5.20e-3  &4.17e-2  &229   \\
2155$-$304    &0.313   &3.10e-2 &2.73e-3 &0.147    &0.166    &1627  \\
2200$+$420    &0.153   &2.64e-2 &8.90e-3 &0.136    &0.752    &332   \\ 
2344$+$514    &7.33e-3 &4.87e-3 &3.83e-3 &7.47e-3  &1.20e-2  &1141  \\ 
2356$-$309    &1.61e-2 &1.30e-2 &2.24e-3 &8.22e-3  &1.15e-3  &1.3e4 \\
\hline
\end{tabular}
\end{table*}

\section{Appendix}

We report here (figures and tables) the SEDs of all blazars in the
sample (Figures 7--13), together with the results of the modelling
(Tables 1 and 2).

\begin{figure*}
\psfig{file=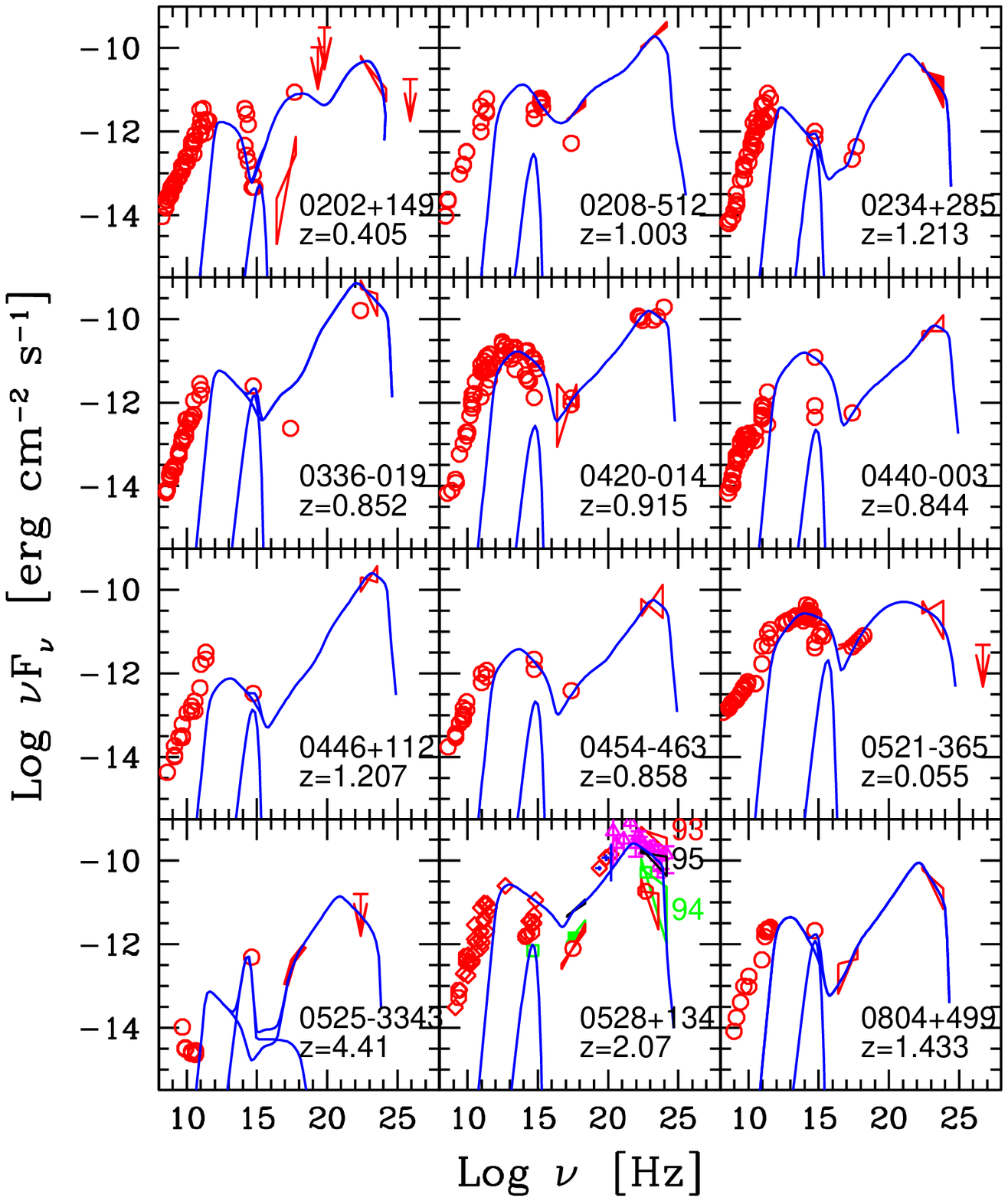,width=22truecm,height=22truecm} 
\vskip -0.7 true cm
\caption{
SEDs of the blazars in our sample. The lines are the result of our
modelling, with the parameters listed in Tab. 1.}
\label{qso1}
\end{figure*}
\begin{figure*}
\psfig{file=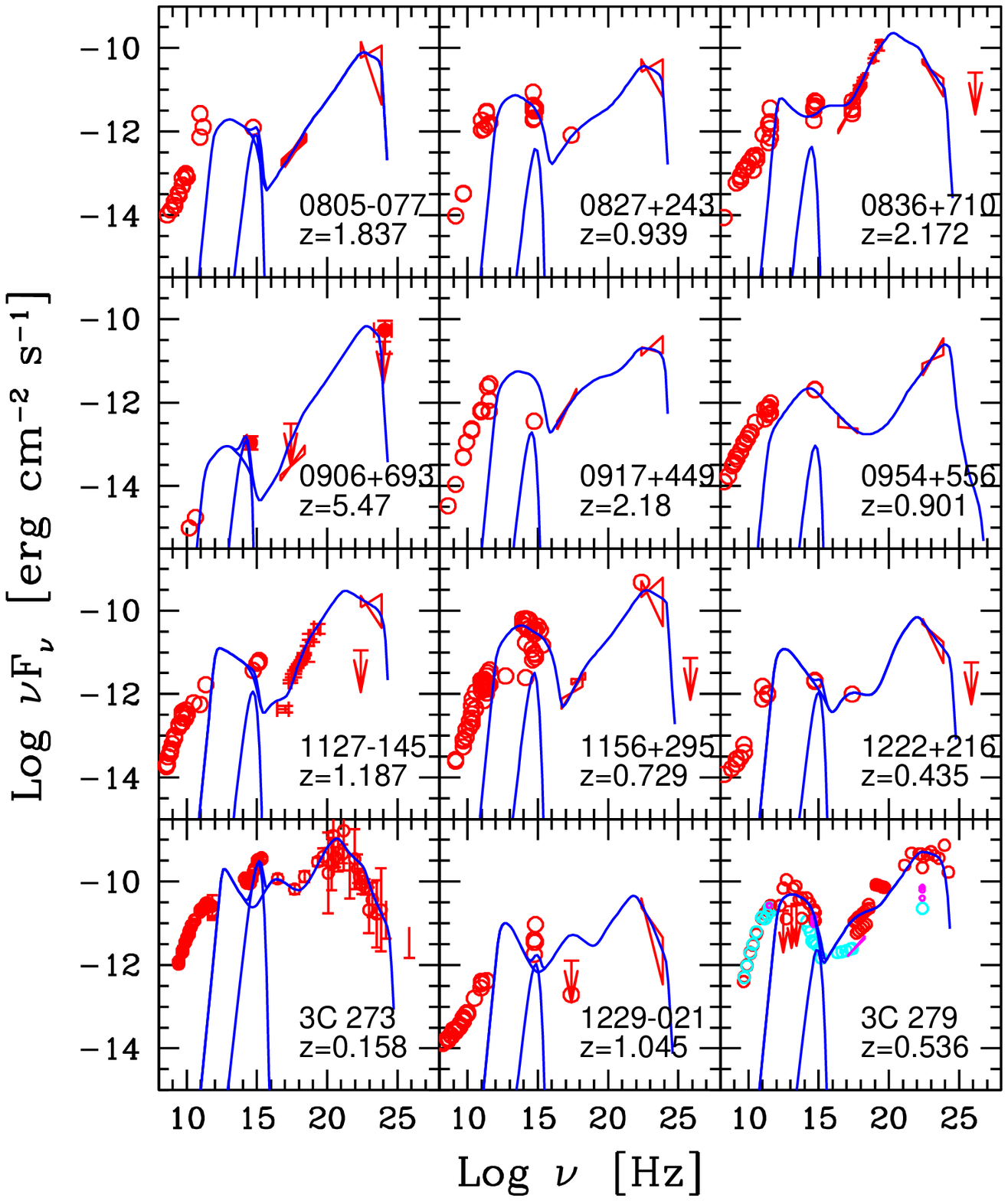,width=22truecm,height=22truecm}
\vskip -0.7 true cm
\caption{
SEDs of the blazars in our sample. The lines are the result of our
modelling, with the parameters listed in Tab. 1.}
\label{qso2}
\end{figure*}
\begin{figure*}
\psfig{file=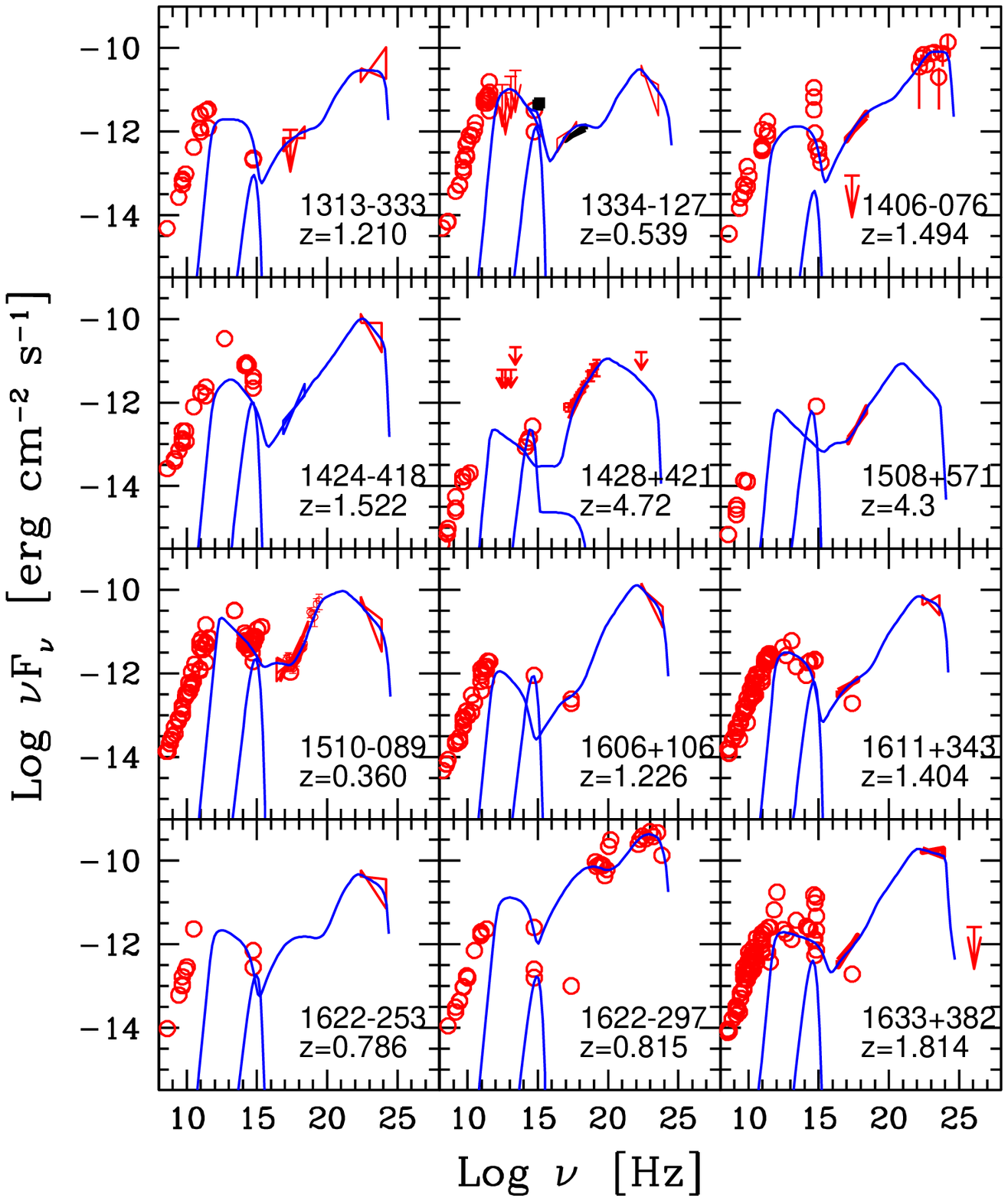,width=22truecm,height=22truecm}
\vskip -0.7 true cm
\caption{
SEDs of the blazars in our sample. The lines are the result of our
modelling, with the parameters listed in Tab. 1.}
\label{qso3}
\end{figure*}
\begin{figure*}
\psfig{file=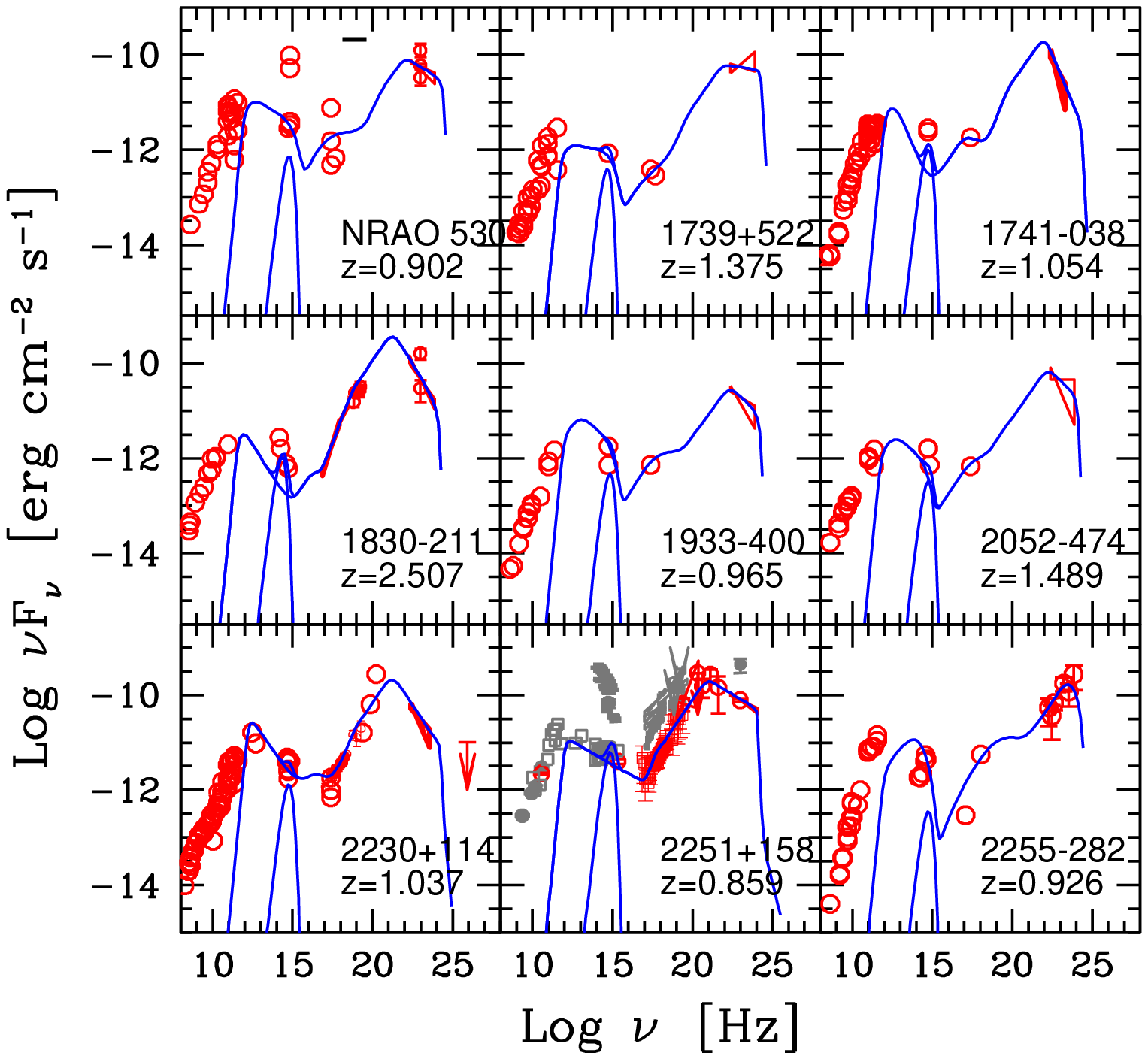,width=22truecm,height=22truecm}
\vskip -0.7 true cm
\caption{
SEDs of the blazars in our sample. The lines are the result of our
modelling, with the parameters listed in Tab. 1.}
\label{qso4}
\end{figure*}
\begin{figure*}
\psfig{file=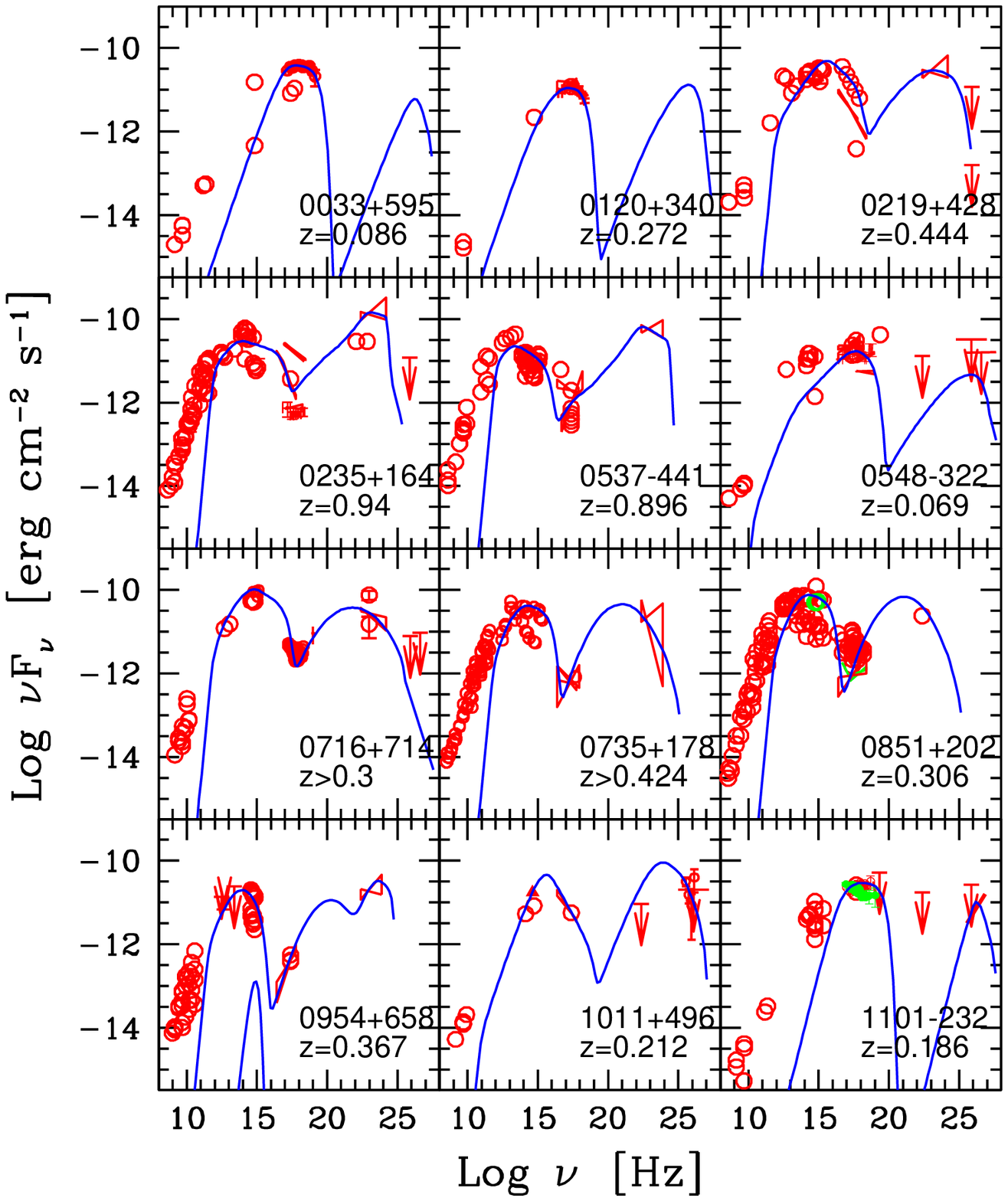,width=21truecm,height=21truecm}
\vskip -0.7 true cm
\caption{
SEDs of the blazars in our sample. The lines are the result of our
modelling, with the parameters listed in Tab. 1.}
\label{bl1}
\end{figure*}
\begin{figure*}
\psfig{file=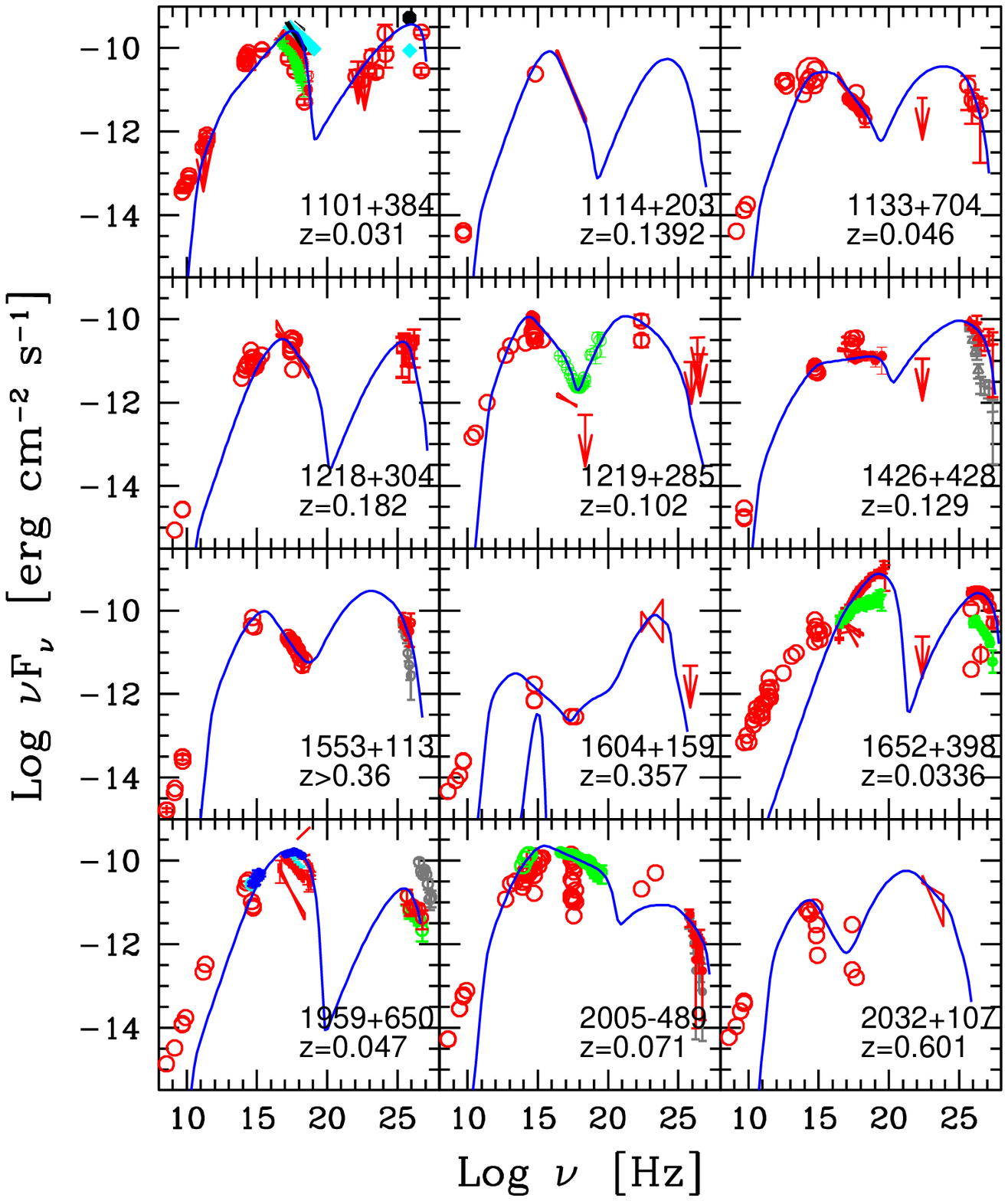,width=21truecm,height=21truecm}
\vskip -0.7 true cm
\caption{
SEDs of the blazars in our sample. The lines are the result of our
modelling, with the parameters listed in Tab. 1.}
\label{bl2}
\end{figure*}
\begin{figure*}
\psfig{file=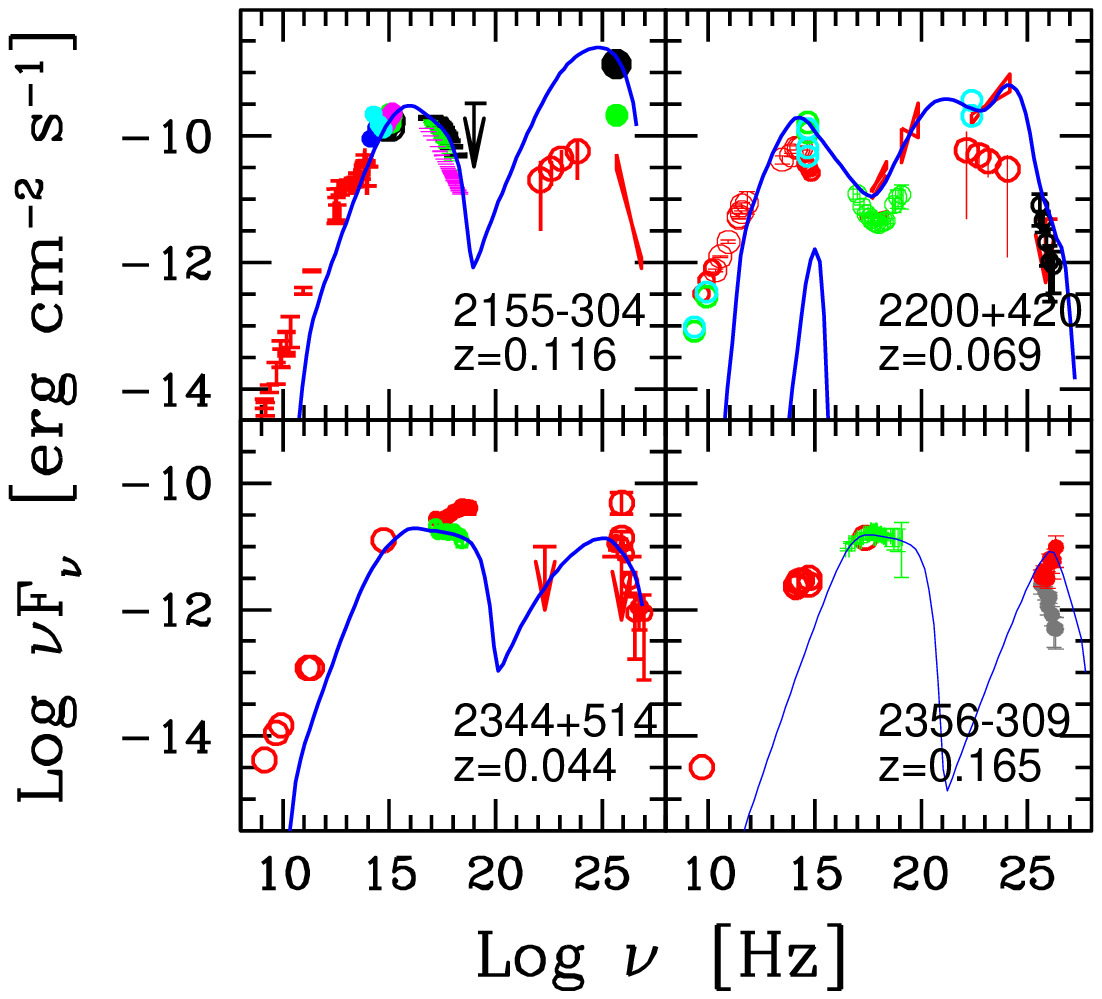,width=21truecm,height=21truecm}
\vskip -9 true cm
\caption{
SEDs of the blazars in our sample. The lines are the result of our
modelling, with the parameters listed in Tab. 1.}
\label{bl3}
\end{figure*}

\end{document}